\documentclass[sigconf]{acmart}

\AtBeginDocument{%
  \providecommand\BibTeX{{%
    \normalfont B\kern-0.5em{\scshape i\kern-0.25em b}\kern-0.8em\TeX}}}

\usepackage{graphicx,acmart-taps}
\usepackage{multirow}
\usepackage{subcaption,colortbl}
\usepackage{color,pifont}
\usepackage{enumitem}
\usepackage{tabularray}
\usepackage{caption}
\usepackage{hyperref}
\usepackage{booktabs}
\usepackage{wasysym}
\usepackage{array}
\usepackage{colortbl}
\usepackage{rotating}

\definecolor{Gallery}{HTML}{D3D3D3}

\definecolor{Alto}{rgb}{0.90,0.90,0.90}

\definecolor{Shark}{rgb}{0.125,0.125,0.133}
\definecolor{MineShaft}{rgb}{0.133,0.133,0.133}
\definecolor{Boulder}{rgb}{0.466,0.466,0.466}

\copyrightyear{2026}
\acmYear{2026}
\setcopyright{cc}
\setcctype{by}
\acmConference[CHI '26]{Proceedings of the 2026 CHI Conference on Human Factors in Computing Systems}{April 13--17, 2026}{Barcelona, Spain}
\acmBooktitle{Proceedings of the 2026 CHI Conference on Human Factors in Computing Systems (CHI '26), April 13--17, 2026, Barcelona, Spain}
\acmPrice{}
\acmDOI{10.1145/3772318.3790383}
\acmISBN{979-8-4007-2278-3/2026/04}

\begin{document}

\title[]{A Systematic Review of User Experiments Measuring the Effects of Dark Patterns}

\author{Brennan Schaffner}
\affiliation{%
 \institution{Georgetown University}
 \city{Washington}
 \state{District of Columbia}
 \country{USA}
}
\email{brennan.schaffner@georgetown.edu}
\author{Luis Heysen}
 \affiliation{%
 \department{Department of Computer Science}
 \institution{University of Chicago}
 \city{Chicago}
 \state{Illinois}
 \country{USA}
}
\email{lheysent@uchicago.edu}
\author{Marshini Chetty}
\affiliation{%
 \department{Department of Computer Science}
 \institution{University of Chicago}
 \city{Chicago}
 \state{Illinois}
 \country{USA}
}
\email{marshini@uchicago.edu}

\begin{abstract}
Deceptive/Manipulative Patterns (DMP) are interface designs, also known as ``dark patterns,'' that manipulate user behavior. While considerable attention has been paid to their ethical and legal implications, empirical evidence about their real-world effects remains diffuse.
This review synthesizes up-to-date experimental studies, focusing on works that quantify how (or whether) DMPs influence users. We also aggregate findings on interventions aimed at reducing DMP effects.
Our synthesis highlights the experimental agreement that DMPs do significantly alter user behavior (with large variance in effect size) and that external interventions have been mostly unsuccessful in mitigating their effects. 
Lastly, we show that significant correlations between DMP effects and personal characteristics (e.g., age or political affiliation) are uncommon, indicating DMPs similarly affected nearly all populations tested.
By summarizing the experimental evidence, we clarify the effects of DMPs, highlight gaps and tensions in the existing experimental literature, and help inform ongoing research and policy directions.
\end{abstract}



\maketitle
\section{Introduction}
\label{sec:introduction}

Deceptive/Manipulative Patterns (DMPs)\footnote{We use Deceptive/Manipulative Patterns (DMPs) to refer to ``dark patterns'' as currently recommended by the Association for Computing Machinery (ACM)~\cite{acm_words_archive}. For the sake of discoverability and connection to prior work, we retain ``dark pattern'' in the title and as a keyword given its  common use by researchers, legislators, and litigators.} are ``user interface design choices that benefit an online service by coercing, steering, or deceiving users into making decisions that, if fully informed and capable of selecting alternatives, they might not make''~\cite{mathur2019dark}. Common examples include platforms employing excessively difficult processes to cancel online subscriptions or change privacy-unfriendly default settings.
DMPs have provoked mounting scrutiny from scholars, regulators, and practitioners, and evidence about their measurable impact on users steadily accumulates~\cite{narayanan2020dark, brenncke2024regulating, fansher2018darkpatterns, 10.1145/3313831.3376321}.
There is general agreement about DMPs' pervasiveness and harmful effects.
This sentiment is supported by several important studies, such as the unifying of foundational taxonomies into a cohesive ontology~\cite{gray2024ontology}, measuring the ubiquity of DMPs~\cite{mathur2019dark, di2020ui}, and studying their potential harms~\cite{mathur2021makes}. 
Yet, skepticism duly remains. It is not well understood whether DMPs have quantified user harms.
Accordingly, from the perspective of industry and marketing practitioners, DMPs are often considered standard practice, reflecting status quo sales strategies that have long been prevalent in commerce~\cite{mathur2021makes, ariely2008predictably}.

The researcher response has been to conduct properly controlled experimentation on the effects of DMPs. There are many challenges to conducting studies that isolate the effects of DMPs threatening ecological validity, including operationalizing DMPs in near real-world environments while considering potential participant harms.
Nevertheless, researchers have designed studies to successfully isolate the effects of DMPs.
For example, an early publication in DMP scholarship conducted a randomized control trial experiment quantifying the effect that different DMPs had on the rates in which consumers stayed enrolled in a malicious subscription~\cite{luguri2021shining}.

Official investigations and court actions have concluded that leaving DMPs unchecked can result in consumer harm. The United States Federal Trade Commission's 2022 report on DMPs describes numerous accounts of manipulative designs extracting millions of dollars from users and tricking them into sharing personal data~\cite{ftc_dpreport_2022}. 
In response, there has been a clear rise in regulatory actions against DMPs~\cite{ftc_dpreport_2022, oecd2022dark, eu_edpb_dps}, sometimes manifesting in high-profile cases. 
For instance, Epic Games agreed to pay a \$245 million dollar settlement for using DMPs to trick children into making unwanted purchases~\cite{ftc_epicgames_2023}.
The accelerating regulatory landscape highlights the role of experimental evidence informing policy.
Although such experiments have entered the landscape, the results are scattered across disciplines and venues, leaving regulators and researchers without a clear, cumulative picture of what is known and where the gaps lie.

The present work quantitatively assists the discussions around whether DMPs change user behavior by synthesizing relevant, to-date experimental evidence supporting (and, to a minor extent, contesting) the harmful effects that DMPs have on users. 
Previous reviews have targeted adjacent yet distinct topics, broadly reviewing designs that influence behavior~\cite{10.1145/3701571.3701572, 10.1145/3461702.3462532}---especially those aimed to \textit{help} users rather than harm them~\cite{10.1145/3054926, HUMMEL201947}---or small subsets of DMPs specifically~\cite{bielova2023survey, 10.1145/3659945, monge2023defining, 10.1145/3368860.3368865}. 
By contrast, our review captures the breadth of all current scholarship that experimentally measures DMP harms.

Building on this synthesis, our review also evaluates the state of DMPs experimental work with four additional analytical angles:
We (i) synthesize results from experiments that have tested potential interventions aimed at combating the effects of DMPs, such as DMP educational sessions or interstitials that promote reflection, (ii) aggregate experiments that have tested marginal effects of ``stacking'' multiple DMPs, (iii) examine which types of DMPs have been tested more frequently and whether specific instances are more effective than others, and (iv) summarize evidence on whether users' personal characteristics mediate the effects of DMPs. 

In sum, our review of experimental DMP literature holds the following objectives: 
\begin{itemize}
    \item In which areas do experimental results agree and/or diverge with respect to quantitative measures of DMPs' effects as well as interventions designed to counteract them? 
    \item What aspects of DMPs have been well described with quantitative measures, and what aspects have been understudied?
\end{itemize}

Considering the increased regulatory efforts aimed at curbing the use of DMPs,\footnote{https://www.deceptive.design/cases} answering these questions is important for aligning the DMPs community and guiding legislative directions.
Our review finds experimental agreement that DMPs commonly have measurable effects on user behavior and that interventions have been thus far ineffective mitigators. 
We also reveal which types of DMPs have been studied most thoroughly and which have been understudied experimentally. 
Ultimately, our review helps researchers and policymakers focus future research and regulatory efforts where they are most needed.

\section{Related Work}
\label{sec:related}
We summarize the present field of DMPs and the relevant previously conducted systematic reviews. 

\subsection{Research On DMPs}
Deceptive and manipulative digital interfaces have become a popular matter of focus across regulatory~\cite{brenncke2024regulating, herman2024dark, oecd2022dark}, academic~\cite{narayanan2020dark, 10.1145/3563703.3596635}, and design spaces~\cite{fansher2018darkpatterns, nngroup_dps}. 
Recent legislation has made efforts to specifically define and ban DMPs in specific contexts.
For example, two of the earliest legal codifications of DMPs are in the 
California Privacy Rights Act in the United States (U.S.) which defines DMPs as ``user interface[s] designed or manipulated with the substantial effect of subverting or impairing user autonomy, decision making, or choice''~\cite{cpra2020}, and the European Union's Digital Services Act which define DMPs as ``[online] practices that materially distort or impair, either on purpose or in effect, the ability of recipients of the service to make autonomous and informed choices''~\cite{dsa2022}. 

Scholars have substantially elaborated upon the harms that DMPs may cause, including harms to individual welfare, collective welfare, and individual autonomy~\cite{mathur2021makes}.
For example, cookie consent banners may bury privacy-friendly options and employ confusing controls resulting in users ``opting'' for more data to be shared with the platform and third-parties. 

Taxonomic efforts have iterated and evolved the classifications of DMPs since they were originally cataloged in 2010~\cite{dporiginal}. Gray et al.'s contemporary ontology published in 2024 consists of five High-Level DMP strategies: Social Engineering, Obstruction, Sneaking, Interface Interference, and Forced Action~\cite{gray2024ontology}. 
Despite their potential for harm, DMPs have been found to be quite prevalent. In 2020, researchers found DMPs in 95\% of 240 popular mobile apps from the Google Play Store~\cite{di2020ui} and in 89\% of cookie banners in the EU~\cite{10.1145/3313831.3376321}.

To measure the effects that DMPs have on users, researchers have begun employing controlled experiments, the earliest of which---published in 2019---studied how 80,000 users of a German website interacted with its consent banner~\cite{10.1145/3319535.3354212}. They found that cookie banner implementation using DMPs resulted in significantly higher consent rates than those without these patterns. 
Experiments have also demonstrated the potential for financial harms from DMPs~\cite{luguri2021shining, zac2023dark, KOH2023100145}. For example, an early DMP experiment from the field of law revealed the power of DMPs in tricking consumers into enrolling in a dubious paid subscription~\cite{luguri2021shining}. 
Our work is the first systematic review aggregating the results of all such experiments---those that measure the effects of DMPs---to date. 

\subsection{Relevant Extant Systematic Reviews of DMPs}
Given the growth in scholarship around DMPs, there has been a corresponding effort to conduct systematic literature reviews to summarize findings and identify new research angles. 
However, no prior systematic review has analyzed the results of all peer-reviewed,  experimental studies across the domain of DMPs. In this section, we discuss the systematic reviews that this work builds upon and highlight the need for the present work. 

\begin{figure*}
    \centering
    \includegraphics[width=0.95\linewidth]{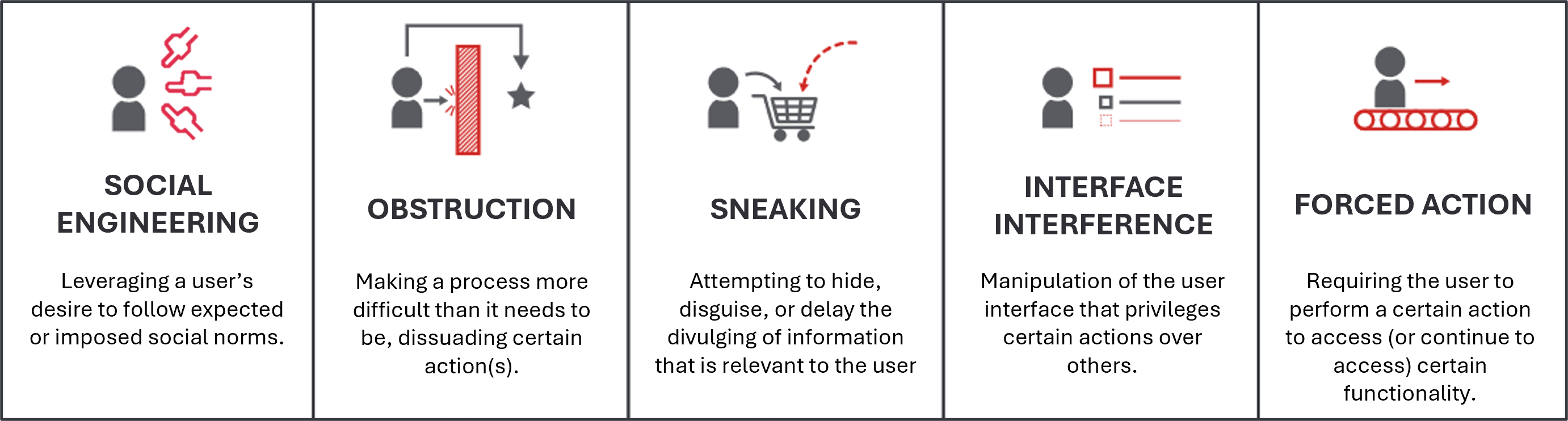}
    \caption{The five high-level categories in Gray et al.'s ontology. Each high-level category also comprises meso- and low-level patterns. This is an edited figure from Gray et al.~\cite{gray2018dark} updated to match the contemporary ontology~\cite{gray2024ontology}.}
    \label{fig:ontology}
    \Description{Small images illustrating the high-level categories in Gray et al.'s ontology accompanied by abridged definitions.}
\end{figure*}

\paragraph{Reviews that Broadly Address DMPs.} 
Multiple works have reviewed and synthesized research on DMPs, aiming to define, classify, and characterize how manipulative design techniques are studied and understood in digital platforms.
In a work-in-progress paper, Gray et al.\ conducted a general review outlining the high-level components of DMPs literature~\cite{10.1145/3563703.3596635}. They found that the most common DMP contributions took the form of \textit{describing} or \textit{framing} DMPs, and the most common methodologies used were content analysis, followed by experiments (for which they identified 14 works at the time). Yi and Li conducted a DMP systematic review focusing on the wide array of regulatory approaches, advocating for a paradigm shift from harm-based frameworks to proactive alternatives~\cite{yi2025mapping}.
Several systematic reviews focused on synthesizing identified DMPs into a cohesive, standard taxonomy~\cite{gray2024ontology, potel2023dark, mathur2021makes}. 
These existing taxonomies informed our approach: specifically, in our methods (see \S\ref{paper_coding}), we use Gray et al.'s ontology to categorize DMPs present in the experimental papers reviewed. See Figure~\ref{fig:ontology} for abridged definitions of the five high-level DMP categories.

Independent and government organizations have also released technical reports including substantial reviews that outline many findings relevant to DMPs (e.g.,~\cite{oecd2022dark}). These reports tend to cover a wide range of information and apply less defined criteria for determining which works to include, resulting in reports that are broadly educational and less reproducible compared to the systematic review of experiments on DMPs conducted in the present work.

\paragraph{Reviews of Influential Design}
There are also a number of related systematic reviews that do not rely on the framing of DMPs but have peripheral overlap in contribution to the DMPs space.
For instance, there are several reviews focusing on the concept of ``nudging''~\cite{10.1145/3054926, HUMMEL201947, ioannou2021privacy}.
Acquisti et al.\ review privacy and security literature related to nudging looking broadly at the limitations and potential of existing interventions and give guidelines for ethical nudge design~\cite{10.1145/3054926}. 
Multiple studies have been conducted that quantitatively reviewed nudging literature to assess the effectiveness of nudging interventions in general and in different application contexts~\cite{HUMMEL201947, ioannou2021privacy}. Both DMPs and nudging could be considered theoretical subsets of persuasive technology~\cite{10.1145/3701571.3701572} or automated influence~\cite{10.1145/3461702.3462532}, both of which have been systematically reviewed for various ethical considerations~\cite{10.1145/3701571.3701572, 10.1145/3461702.3462532}.

In our review, we focus specifically on DMPs literature, which we distinguish from nudging following previous literature (e.g., ~\cite{10.1145/3476087}), where nudges are ``soft paternalism'' that guide users towards better decisions~\cite{10.1145/3054926, 5370707} and dark pattens harm a user's ability to make informed decisions~\cite{gray2018dark, gray2024ontology, mathur2021makes}. In short, while both are operationalized by changes to choice architecture, nudges are said to help users, whereas DMPs are said to harm users.

\paragraph{Context-Specific Reviews On DMPs.}
Some reviews have focused on specific subsets or contexts of DMPs. 
For instance, Bielova conducted a review of DMP effects on users' acceptance rates in cookie banners with the French Data Protection Authority~\cite{bielova2023survey}. 
Hadan et al.\ reviewed the DMP literature in specifically extended reality (XR) environments identifying risks and harms unique to XR~\cite{10.1145/3659945}. 
Westin and Chiasson review DMPs literature through a paradigm of social pressures~\cite{10.1145/3368860.3368865}. 
In another review, Roffarello et al.\ develop a framework for a specific subset of DMPs related to attentional harms, defining ``attention capture damaging patterns''~\cite{monge2023defining}. Our review captures multiple domains of study and instead focuses on experiments quantifying the effects of various DMPs.

\paragraph{Reviews On DMP Harms.}
Some reviews focus on the harms of DMPs~\cite{santos2025no, ahuja2022conceptualizations, mathur2021makes}. For instance, Mathur et al.\ review DMPs literature and outline normative perspectives through which the harms of DMPs can be analyzed~\cite{mathur2021makes}.
Sanju and Kumar align the types of DMPs with specifically how they harm user autonomy along four dimensions: agency, freedom of choice, control, and independence~\cite{ahuja2022conceptualizations}.
In another review, Cara sorted DMPs by their harm severity, from ``just annoying'' to ``need official regulation''~\cite{cara2019dark}.
Most recently, Santos et al.\ review relevant DMPs literature to create a taxonomy of the possible harms that DMPs can cause and discuss the challenges surrounding the more complicated harms such as those that are non-material or societal~\cite{santos2025no}. 
In our review, we build on this literature to systematically look at studies that have \textit{experimentally measured} these harms.

\section{Review Scope}
\label{sec:scope}

Here we outline the scope of our systematic review. We outline our inclusion parameters for the studies we ultimately analyzed as part of our dataset.

\paragraph{DMP Framing.} The study must use the framing of DMPs, meaning the paper would have to contain the phrase ``DMP''---or its parallel terms (i.e., ``deceptive and manipulative design'' and ``deceptive design patterns''---in the work's framing. Effectively, this means the terminology had to appear in the title, abstract, keywords, or background sections to be included in our dataset.

\paragraph{Peer-Reviewed.}
We excluded studies that were not peer-reviewed, including technical reports, academic theses, and white papers.
However, we included preprints that had gone through the peer-review process and were yet to appear at a pending conference. The inclusion of government reports (as discussed in the \S\ref{sec:methodology:corpus_creation}) is the only exception to this criterion or any other criterion in this section. 

\paragraph{User Experience (UX) Elements That Lead To Harm.} To distinguish from literature that uses persuasive design largely for the user's benefit, we excluded studies that do not engage with known user harms. In practice, this exclusion separates the work most notably from those of ``Nudge Theory,'' which is primarily housed in the behavioral sciences and originated with the intention of paternal aid for user decision making, especially according to those who popularized it~\cite{thaler2021nudge}. Some researchers consider DMPs as an (evil) type of nudge. We do not exclude such studies since they retain the framing and harmful character of DMPs. 
This criterion also led to the exclusion of papers that measure ``neutral'' persuasive design, where the platform benefit and user harms are unclear (e.g.,~\cite{10.1145/3640543.3645202}).

\paragraph{Direct Measures of Behavior.}
In spirit of building the strongest suite of evidence, we excluded works that rely on self-reported metrics instead of direct measures of participant behavior. In practice, this marks the difference between what participants \textit{actually do} when faced with DMPs and what they \textit{say they would do} when presented with an example of a DMP in a survey. 
This also led to the exclusion of studies that used survey constructs or scales as experimental dependent variables (e.g., perceived platform trustworthiness) because they are indirect proxies for behavior rather than direct behavioral measures.

\paragraph{Dependent Variables Directly Relevant to DMPs}
We excluded studies that examined the effects of DMPs on behavior that are tangential to the ``intent'' of the DMP. For example, some studies measure how DMPs increase users' ``consent'' to the sale of their personal data (a direct effect) as well as how the DMPs may affect how long it takes users to make the consent decision (an indirect effect). 
Studies about the indirect effects are excluded for the sake of simplifying the interpretation of the findings. 

\paragraph{Statistical Significance on Quantitative Metrics}
Due to industry responses to both enacted regulation and threats thereof that DMPs are not well-studied or definitively harmful to consumers, we chose to seek what is most popularly accepted to be ``hard'' evidence. That is, studies must test for statistical significance as well as report corresponding statistical measures (e.g., p-values) in order to be included in the analysis.

Unfortunately, these criteria led to the exclusion of papers that quantitatively or descriptively measured how DMPs affected behavior---often to a convincing extent---but did not calculate statistical significance. For example, Monge Roffarello and De Russis qualitatively found that removing certain DMPs on social media led to reductions in time spent on the platform~\cite{monge2022towards}.

The dataset ultimately included the scientific methodologies typically deemed most rigorous (i.e., ``true experiments''), including between-subjects post-test designs with control groups, within-subjects repeated measures designs with counterbalancing, 
or a mix of both.

Such controlled and statistically-minded behavioral experiments included in the resulting dataset are not always practical, necessary, or superior in demonstrating the effects of deceptive and manipulative design~\cite{willis2020deception}.
Our choice to steel man the position of a DMP skeptic should not be interpreted as a value or relevancy judgment on the plethora of impressive and important studies that measure the effects of DMPs qualitatively or through expert or legal analysis. See \S\ref{sec:evidence} for further discussion on this matter. 

Lastly, many works use a combination of included and excluded studies together in any single publication. We include only the relevant components from each work. 
When multiple works were found to include duplicately-published content, relevant results were only included once.

\section{Methodology}
Here we present the processes employed to collect the corpus of experiments fitting our review criteria and how they were analyzed.

\begin{figure}[htbp]
    \centering
    \includegraphics[width=0.95\linewidth]{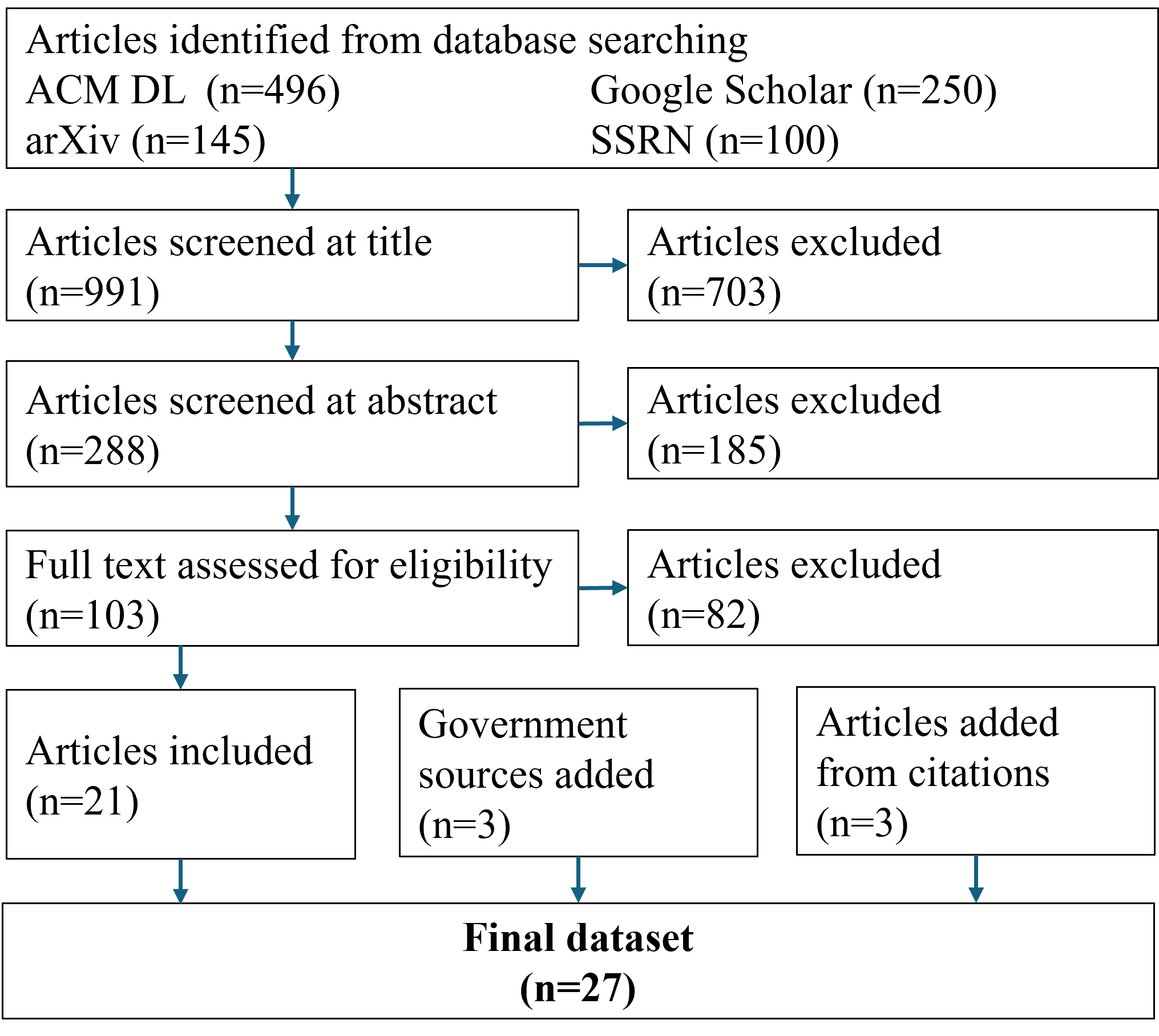}
    \caption{Flow of study selection through the screening process, from 991 initial papers filtered down to the final 27.}
    \label{fig:screening-graph}
    \Description{A boxes-and-arrows diagram showing the filtering of papers from top to bottom, starting with the initial 991 papers down to the final 27. The filtering stages include title screening, abstract screening, then full text screening. Three government studies are added a the end, as well as three studies discovered by consulting the citation lists of the otherwise included studies.}
\end{figure}

\subsection{Corpus Creation}
\label{sec:methodology:corpus_creation}

We assembled a comprehensive corpus of experimental literature on DMPs by systematically searching multiple academic repositories between January and May 2025.\footnote{Search queries are provided in Appendix~\ref{sec:queries}.} Our repository selection ensured coverage across technical, legal, and social science fields.

For breadth across disciplines, we employed a Python-based Selenium web scraper to retrieve the first 250 results from \textbf{Google Scholar}, in line with prior work~\cite{10.1145/3563703.3596635}.\footnote{\href{https://scholar.google.com/}{https://scholar.google.com/}} 
For computer science publications, our search returned 496 results in the \textbf{ACM Digital Library},\footnote{\href{https://dl.acm.org/}{https://dl.acm.org/}} which we retrieved using the platform's native batch download feature.
To capture works in the adjacent fields of law and social science research, we downloaded the first 100 results from \textbf{SSRN},\footnote{\href{https://www.ssrn.com/index.cfm/en/}{https://www.ssrn.com/index.cfm/en/}} reflecting its more specialized scope and tangential relevance to Human-Computer Interaction. 
Lastly, we obtained the 145 publications our query returned in \textbf{arXiv}\footnote{\href{https://arxiv.org/}{https://arxiv.org/}} to ensure technical coverage and capture to-be-published preprints.

\aptLtoX{
    \begin{table}
    \caption{The publication domain, venue, and year for the 27 final papers in the dataset. The domain column also shows the distribution of experimental units across domains in parentheses.}
    \label{tab:domainvenueyear}
    \centering
    \begin{tabular}{|l c|l c|l c|}
    \hline
    \multicolumn{2}{|c}{\textbf{Domain}} &
    \multicolumn{2}{c}{\textbf{Venue}} &
    \multicolumn{2}{c|}{\textbf{Year}} \\
    \hline
    Consent Popups   & 16 (51) & Conference       & 12 & 2025 & 4 \\
    Subscriptions    & 4 (41)  & Journal          & 8  & 2024 & 4 \\
    Privacy Settings & 2 (27)  & Non-Archival     & 3  & 2023 & 6 \\
    E-Commerce       & 2 (24)  & Government Study & 3  & 2022 & 4 \\
    Streaming        & 1 (3)   & Preprint         & 1  & 2021 & 6 \\
    Advertising      & 1 (1)   &                 &    & 2020 & 2 \\
    Donations        & 1 (1)   &                 &    & 2019 & 1 \\
    \hline
    \end{tabular}
    \end{table}

    \begin{figure}
    \centering
            \includegraphics[width=\linewidth]{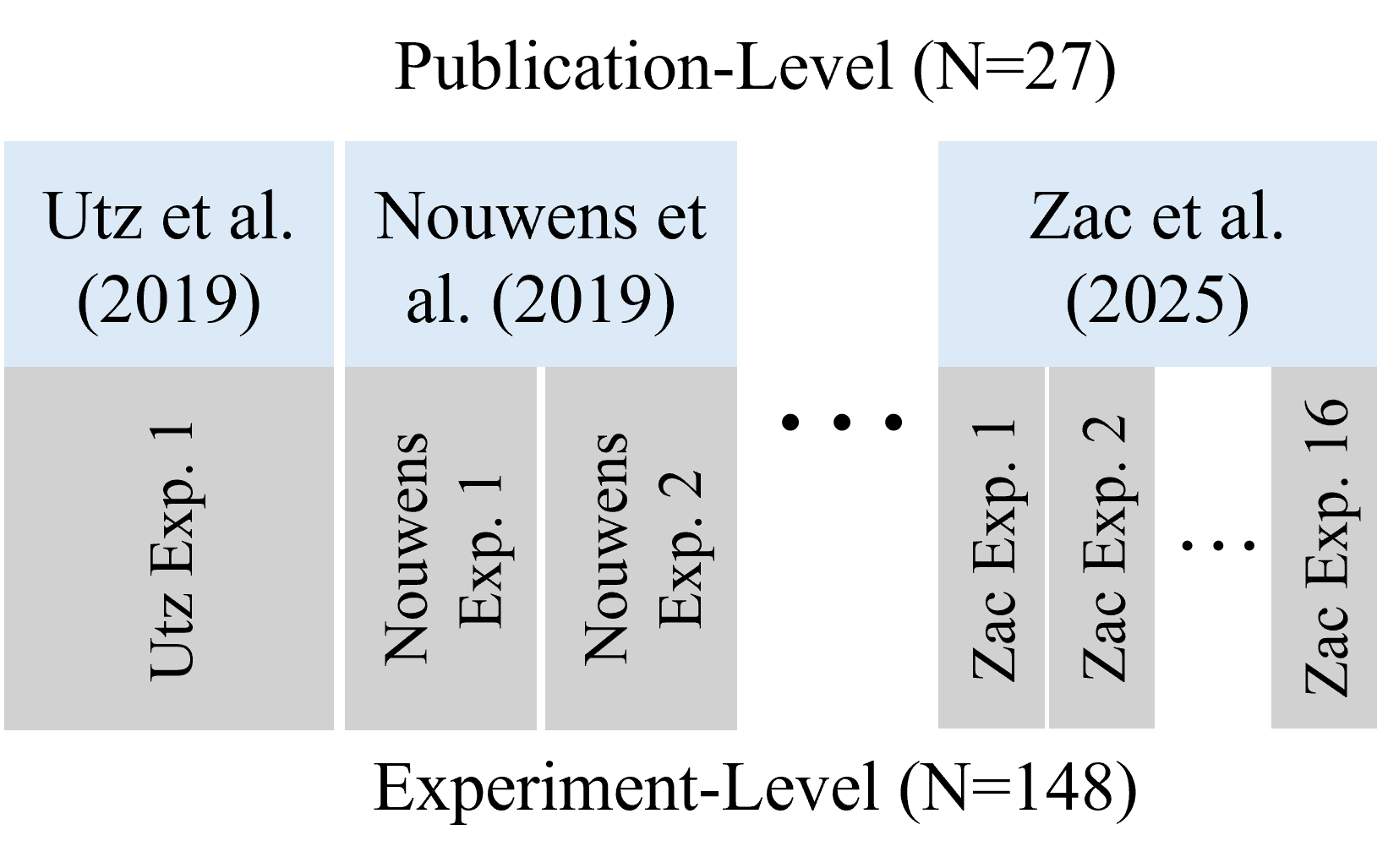}
            \captionof{figure}{Visual depiction of both the publication- and experiment-levels in the dataset.}
            \label{fig:publevel_explevel}
            \Description{A visual illustrating how the 27 final publications are further subdivided by the number of experimental units in each. It shows, for sake of example, that Utz et al contains 1 experiment, Nouwens et al contains 2, and Zac et al contains 16. From hereon, the paper uses both publication-level to refer to the 27 papers and experiment-level to refer to the 148 experiments conducted within those 27 papers.}
        \end{figure}
    }{
    \begin{figure*}[htbp]
    \begin{minipage}{\textwidth}
        \begin{minipage}[]{0.48\textwidth}
                \captionof{table}{The publication domain, venue, and year for the 27 final papers in the dataset. The domain column also shows the distribution of experimental units across domains in parentheses. }           
                \label{tab:domainvenueyear}
                \resizebox{\linewidth}{!}{ 
                    \begin{tblr}{
                    row{1} = {c},
                    cell{1}{1} = {c=2}{},
                    cell{1}{3} = {c=2}{},
                    cell{1}{5} = {c=2}{},
                    vline{2,4} = {1}{},
                    vline{3,5} = {2-8}{},
                    hline{1-2,9} = {-}{},
                    }
                    \textbf{Domain } &         & \textbf{Venue }  &    & \textbf{Year } &   \\
                    Consent Popups   & 16 (51) & Conference       & 12 & 2025           & 4 \\
                    Subscriptions    & 4 (41)  & Journal          & 8  & 2024           & 4 \\
                    Privacy Settings & 2 (27)  & Non-Archival     & 3  & 2023           & 6 \\
                    E-Commerce       & 2 (24)  & Government Study & 3  & 2022           & 4 \\
                    Streaming        & 1 (3)   & Preprint         & 1  & 2021           & 6 \\
                    Advertising      & 1 (1)   &                  &    & 2020           & 2 \\
                    Donations        & 1 (1)   &                  &    & 2019           & 1 
                    \end{tblr}

                }
        \end{minipage}
        \hfill
        \begin{minipage}[]{0.48\textwidth}
            \includegraphics[width=\linewidth]{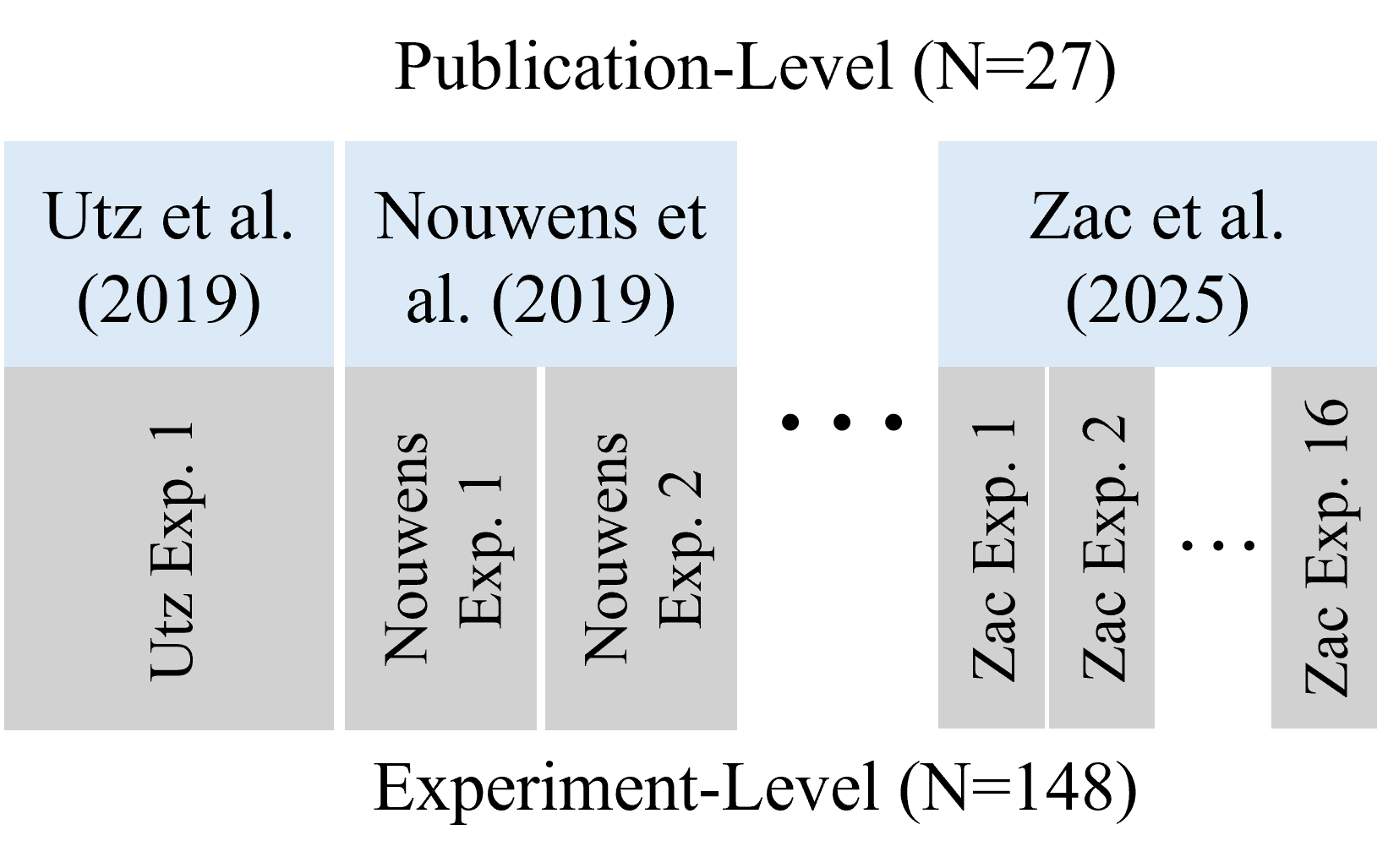}
            \captionof{figure}{Visual depiction of both the publication- and experiment-levels in the dataset.}
            \label{fig:publevel_explevel}
            \Description{A visual illustrating how the 27 final publications are further subdivided by the number of experimental units in each. It shows, for sake of example, that Utz et al contains 1 experiment, Nouwens et al contains 2, and Zac et al contains 16. From hereon, the paper uses both publication-level to refer to the 27 papers and experiment-level to refer to the 148 experiments conducted within those 27 papers.}

        \end{minipage}
    \end{minipage}
\end{figure*}}

All search results were exported as BibTeX files and consolidated using Rayyan.ai,\footnote{\href{https://rayyan.ai/}{https://rayyan.ai/}} a collaborative tool for systematic reviews and automated detection of duplicate papers.
In total, our initial dataset contained 991 papers before screening.

\subsubsection{Screening Process}

We applied a structured, multi-stage screening process to identify only the relevant papers. At every stage of screening, we favored inclusion in cases of uncertainty. The process, summarized in Figure~\ref{fig:screening-graph}, consisted of the following steps:

\begin{enumerate}
\item \textbf{Title/Abstract Screening}: From the initial dataset of 991 papers, titles and abstracts were reviewed, and papers clearly not relevant to DMPs or experimental approaches were excluded, resulting in 103 papers. Duplicate papers were also excluded. The most common reasons for exclusion at this stage were irrelevance arising from overlap in search terms with popular concepts in other fields (e.g., `dark matter'). 
\item \textbf{Full-Text Screening}: Full texts were assessed according to the inclusion and exclusion criteria specified in \S\ref{sec:scope}. Three papers had to be retrieved directly from the authors but were ultimately excluded for not meeting the scope criteria. Papers unavailable in English were also excluded. 
The most common reasons for exclusion at this stage were: not conducting controlled experiments, not being peer-reviewed, or measuring user perceptions rather than behavior change.
An accurate count of exclusion reasons is unavailable since analysis was stopped upon encountering the first instance of disqualifying criteria, so not every disqualifying characteristic was recorded. 
\item \textbf{Government Reports}: Based on domain expertise and expert consultation, three relevant government studies meeting the inclusion criteria were incorporated (\cite{lupianez2022behavioural, bogliacino2024testing, sernac}).\footnote{Duplicate findings present in both Bogliacino and EU Commission (2022) are counted only once, attributed to the latter.} 
\item \textbf{Citation Search}: Reference lists of included papers were analyzed, with identified studies subjected to the same screening process, resulting in three additional papers. \end{enumerate}

A total of 27 papers were included in the final dataset.

\subsection{Corpus Analysis}
\label{paper_coding}
We first recorded relevant meta details for each of the publications in the final dataset. 
For each publication in the corpus, we noted the domain (e.g., e-shopping or consent pop-ups), the venue type (e.g., peer-reviewed journal), and publication year, which can be seen in Table~\ref{tab:domainvenueyear}.
Since each publication could include more than one experiment, we then developed a codebook to apply at the \textit{experiment}-level as discussed below.

\subsubsection{Identifying Experimental Units}
To enable consistency across the dataset, we broke down multi-part studies into relevant \textit{experimental units} (illustrated in Figure~\ref{fig:publevel_explevel}), where each \textit{experimental unit} represents a full study fitting the inclusion criteria outlined in \S\ref{sec:scope}. For example, Gra{\ss}l et al.~\cite{grassl21dark} test the difference between several consent dialogues with permuted design against a neutral control dialogue. They test for statistically significant effects between participants in three DMPs conditions and a control condition.  Hence, for the sake of our review, Gra{\ss}l et al.'s work is reduced to three experimental units, one for each DMP effect reported. They also tested for effects of various conditions that ``nudge towards privacy-friendly options'' as well as how various conditions---both DMPs and nudging---affect participants' self reported levels of control and deliberation, all of which are out of scope of this review.

\aptLtoX{
  \begin{table*}
  \centering
  \caption{Summary statistics for the final dataset of 148 experimental units across 27 papers, including the recruitment methods used, the number of experimental units per paper, and the number of participants per experimental unit. Studies could use multiple recruitment methods. Note that the sample sizes reported are skewed higher since many papers reported only a total sample size without reporting the relevant subset sample sizes of each experimental conditions. Posner et al. conducted a large-scale natural experiment; therefore, sample size metrics are reported both including and excluding it (the latter shown in parentheses).}
  \label{tab:dataset_summary}

  \begin{tabular}{lr|l r|l r}
  \hline
  \multicolumn{1}{c}{\textbf{Recruiting Method}} &
  \multicolumn{1}{c}{\textbf{Exp.\ units /}} &
  \multicolumn{2}{c}{\textbf{Exp.\ units}} &
  \multicolumn{2}{c}{\textbf{Sample Size Per Exp.}} \\
  \multicolumn{1}{c}{} &
  \multicolumn{1}{c}{\textbf{No.\ Papers}} &
  \multicolumn{2}{c}{\textbf{Per Paper}} &
  \multicolumn{2}{c}{(without Posner et al.)} \\
  \hline

  Online recruiting firms &
  115 / 15 &
  Median   & 3     &
  Median   & 925 (922) \\

  University connections or mailing lists &
  9 / 4 &
  Average  & 5.48  &
  Average  & 19,314 (1,758) \\

  Live A/B tests &
  9 / 4
  &
  SD       & 6.46  &
  SD       & 212,947 (6,432) \\

 \textit{Not specified} &
  9 / 2 &
  Range    & 1--26 &
  Range    & 40--2.6M (40--70,208) \\

  Google ads &
  9 / 2 &
  &  &
  &  \\

  Other &
  3 / 2 &
  &  &
  &  \\
  \hline
  \end{tabular}
  \end{table*}
}{
  \begin{table*}[bth!]
  \centering
  \caption{Summary statistics for the final dataset of 148 experimental units across 27 papers, including the recruitment methods used, the number of experimental units per paper, and the number of participants per experimental unit. Studies could use multiple recruitment methods. Note that the sample sizes reported are skewed higher since many papers reported only a total sample size without reporting the relevant subset sample sizes of each experimental conditions. 
  Posner et al.\ conducted a large-scale natural experiment; therefore, sample size metrics are reported both including and excluding it (the latter shown in parentheses).}
  \label{tab:dataset_summary}
  \begin{tblr}{
    width=0.8\linewidth,
    row{1} = {c},
    cell{1}{1} = {m},
    cell{1}{3} = {c=2}{},
    cell{1}{5} = {c=2}{},
    cell{2}{2} = {r},
    cell{2}{4} = {r},
    cell{2}{6} = {r},
    cell{3}{1} = {r=2}{},
    cell{3}{2} = {r=2}{r},
    cell{3}{4} = {r},
    cell{3}{6} = {r},
    cell{4}{4} = {r},
    cell{4}{6} = {r},
    cell{5}{2} = {r},
    cell{5}{4} = {r},
    cell{5}{6} = {r},
    cell{6}{2} = {r},
    cell{6}{4} = {r},
    cell{6}{6} = {r},
    cell{7}{2} = {r},
    cell{7}{4} = {r},
    cell{7}{6} = {r},
    cell{8}{2} = {r},
    vline{3,5} = {2-8}{},
    vline{5} = {4}{},
    hline{1-2,9} = {-}{},
  }
  {\textbf{}\\\textbf{Recruiting Method}}     & {\textbf{Exp. units /}\\\textbf{No. Papers}} & {\textbf{Exp. units }\\\textbf{Per Paper}} &        & {\textbf{Sample Size Per Exp.}\\(without Posner et al.)} &                     \\
  Online recruiting firms                     & 115 / 15                                     & Median                                     & 3      & Median                                                        & 925 (922)           \\
  {University connections \\or mailing lists} & 9 / 4                                        & Average                                    & 5.48   & Average                                                       &  19,314 (1758)      \\
                                              &                                              & SD                                         & 6.46   & SD                                                            & 212,947 (6432)    \\
  Live A/B tests                                  & 9 / 4                                        & Range                                      & 1 – 26 & Range                                                         & 40–-2.6M (40--70,208) \\
  \textit{Not specified}                      & 9 / 2                                        &                                            &        &                                                               &                     \\
  Google ads                              & 9 / 2                                        &                                            &        &                                                               &                     \\
  Other                                       & 3 / 2                                        &                                            &        &                                                               &                     
  \end{tblr}
\end{table*}}

For each experimental unit, we noted the following: 
\begin{itemize}
    \item \textbf{DMP(s) Tested.} We used the leading ontology of DMPs from Gray et al.~\cite{gray2024ontology}\footnote{Gray et al.'s taxonomy aggregates terminology from the ten most frequently cited academic and regulatory reports on DMPs.} to standardize the terminology across the corpus. In some cases, this required reclassifying the DMPs reported by the original authors into standard terminology. Moreover, if the authors were testing the effects of a design that did not fit into the ontology, it was excluded for not having scholarly consensus on being a DMP. 
    \item \textbf{Experimental Results.} We coded whether the experiment resulted in statistically significant results, to what degree of significance, and the effect size if reported in universal terms. It is important to note that lacking statistical significance (failing to reject the null hypothesis) is not identical to proving no effect exists (affirming the null hypothesis). 
    \item \textbf{Participant and Recruitment Details.} We recorded the sample size of each experimental unit, recruitment methods, devices used, whether specific demographics were targeted, and study locations. 
    \item \textbf{Other Experimental Details.} We recorded the independent variables (e.g., presence of a DMP), dependent variable (e.g., data shared or product purchased), experiment type (e.g., posttest-only control group design), and experimental setting (e.g., in the wild or online study with deception). 

\end{itemize}

All experimental units fell into one of the following five categories which are discussed separately in the following sections: 

\begin{enumerate}[label={}, leftmargin=0.4cm]
    \item \textbf{Category 1:} tested for effects induced by presence of DMP(s) compared to a control without the DMP(s).
    \item \textbf{Category 2:} tested whether external interventions mitigated the effects induced by DMPs, where external interventions are attempts to reduce DMP effects without removing them altogether.\footnote{The removal of a DMP altogether constitutes a control group in a Category 1 experiment. What distinguishes external interventions is that they represent external systems or changes to the interface that are not directly changing the DMPs. Similarly, ``bright patterns'' are not considered external interventions since they operate within the same vein as the DMPs, essentially removing them altogether.}
    \item \textbf{Category 3:} tested for additive effects when multiple DMPs were present. 
    \item \textbf{Category 4:} compared the effects from distinct types of DMPs. 
    \item \textbf{Category 5:} tested for correlations between the effects of DMPs and user personal characteristics, ranging from demographics (e.g., age and gender) to personal attributes (e.g., technology affinity and political affiliation). These tests were commonly post-hoc analyses at a paper-level (rather than the experiment-level), and therefore these results are excluded from any experiment unit counts in the following sections. We instead present a summarization at the paper-level in \S\ref{findings:personalchar}.
\end{enumerate}

\subsection{Methodological Limitations}
\label{sec:limitations}

Our systematic review methodology has several limitations that should be acknowledged. First, as a snapshot of literature collected up to May 2025, we cannot guarantee inclusion of papers published after our search period. Second, despite our comprehensive search strategy across multiple repositories, we may have missed rare papers not indexed in our selected databases or those using terminology outside our search parameters. We mitigated this limitation by consulting the cited works of the papers resulting from our search. Third, our review may be affected by publication bias, as studies with null results are less likely to be submitted and published, a well-known metascientific phenomenon~\cite{joober2012publication}.
Fourth, the chosen representation of experimental units treats experiments on equal grounds despite their individual differences in important parameters such as sample size. However, this bias is moderately counteracted by larger studies typically conducting more experiments and are thus weighted higher in analytical representation.
Fifth, our search strategy employed cutoff points when reviewing ranked search results for two of the databases (Google Scholar and SSRN), introducing potential for missed relevant works. This risk was mitigated through conducting searches across multiple databases, setting search result thresholds beyond drop-offs in search result relevance, and consulting the citations of the included works for additional relevant studies to include. 
We took a conservative approach to minimize the potential of missing relevant works as evidenced by the high exclusion rate across screening stages.
Last, the inclusion of only English-language publications may have excluded relevant work from non-English speaking regions, potentially limiting the diversity of perspectives in our review.

\aptLtoX{\begin{table*}
\centering
\caption{Breakdown of the the different types of DMPs tested across the corpus at the experiment-level using Gray et al.'s ontology. Experimental units often implemented more than one DMP, hence the column totals surpassing the total experiment count. The numbers in this table represent the total unique instances of each DMP tested in the dataset.}
\label{tab:dp_breakdown}

\resizebox{0.8\linewidth}{!}{
\begin{tabular}{l rl rl r}
\hline
\textbf{High-Level} &  &
\textbf{Meso-Level} &  &
\textbf{Low-Level} &  \\
\hline
\rowcolor{Alto}
\textbf{Interface Interference} & 157 &  &  &  &  \\

 &  & Manipulating Choice Architecture & 107 & Visual Prominence & 61 \\
\rowcolor{Alto}
 &  &  &  & False Hierarchy & 46 \\

 &  & Bad Defaults & 28 &  &  \\
\rowcolor{Alto}
 &  & Emotional or Sensory Manipulation & 13 & Positive or Negative Framing & 13 \\

 &  & Trick Questions & 9 &  &  \\
\rowcolor{Alto}
\textbf{Social Engineering} & 76 &  &  &  &  \\

 &  & Shaming & 28 & Confirmshaming & 28 \\
\rowcolor{Alto}
 &  & Urgency & 19 & Countdown Timer & 9 \\

 &  &  &  & Limited Time Message & 5 \\
\rowcolor{Alto}
 &  &  &  & Activity Message & 5 \\

 &  & Personalization & 11 &  &  \\
\rowcolor{Alto}
 &  & Scarcity and Popularity Claims & 9 & High Demand & 9 \\

 &  & Social Proof & 9 & Low Stock & 9 \\
\rowcolor{Alto}
\textbf{Obstruction} & 45 &  &  &  &  \\

 &  & Adding Steps & 45 &  -- & 40 \\
\rowcolor{Alto}
 &  &  &  & Privacy Maze & 5 \\

\textbf{Forced Action} & 21 & --  & 1  &  &  \\

\rowcolor{Alto}
 &  & Nagging & 15 &  &  \\

 &  & Attention Capture & 3 & Autoplay & 3 \\
\rowcolor{Alto}
 &  & Forced Communication or Disclosure & 2 &  &  \\

\textbf{Sneaking} & 10 &  &  &  &  \\

\rowcolor{Alto}
 &  & Hiding Information & 9 &  &  \\

 &  & Bait and Switch & 1 & Disguised Ads & 1 \\
\hline
\end{tabular}
}
\end{table*}}{
\definecolor{Gallery}{rgb}{0.90,0.90,0.90}
\definecolor{White}{rgb}{1, 1, 1}
\begin{table*}[thbp!]
\caption{Breakdown of the the different types of DMPs tested across the corpus at the experiment-level using Gray et al.'s ontology. Experimental units often implemented more than one DMP, hence the  column totals surpassing the total experiment count. The numbers in this table represent the total unique instances of each DMP tested in the dataset.}
\label{tab:dp_breakdown}
\resizebox{0.8\linewidth}{!}{
\centering
\begin{tblr}{
  row{even} = {Gallery},
  hline{1,26} = {-}{0.08em},
  hline{2} = {-}{},
}
\textbf{High-Level}                         &    & \textbf{Meso-Level}                   &   & \textbf{Low-Level}           &    \\
{\textbf{Interface} \textbf{Interference}} & 157 &                                       &     &                              &    \\
                                            &     & {Manipulating Choice  Architecture}  & 107 & Visual Prominence            & 61 \\
                                            &     &                                       &     & False Hierarchy              & 46 \\
                                            &     & Bad Defaults                          & 28  &                              &    \\
                                            &     & {Emotional or Sensory Manipulation}  & 13  & Positive or Negative Framing & 13 \\
                                            &     & Trick Questions                       & 9   &                              &    \\
{\textbf{Social } \textbf{Engineering}}    & 76  &                                       &     &                              &    \\
                                            &     & Shaming                               & 28  & Confirmshaming               & 28 \\
                                            &     & Urgency                               & 19  & Countdown Timer              & 9  \\
                                            &     &                                       &     & Limited Time Message         & 5  \\
                                            &     &                                       &     & Activity Message             & 5  \\
                                            & & Personalization & 11 & & \\
                                            &     & {Scarcity and Popularity  Claims}    & 9   & High Demand                  & 9  \\
                                            &     & Social Proof                          & 9   & Low Stock                    & 9  \\
\textbf{Obstruction}                        & 45  &                                       &     &                              &    \\
                                            &     & Adding Steps                          & 45  & ~ ~ ~ ~ ~ ~ ~ ~ ~ ~ ~ -~     & 40 \\
                                            &     &                                       &     & Privacy Maze                 & 5  \\
{\textbf{Forced} \textbf{Action}}          & 21  &   ~-~                                    &  1   &                              &    \\
                                            &     & Nagging                               & 15  &                              &    \\
                                            &     & Attention Capture                     & 3   & Autoplay                     & 3  \\
                                            &     & {Forced Communication or Disclosure} & 2   &                              &    \\
\textbf{Sneaking}                           & 10   &                                       &     &                              &    \\
                                            &     & Hiding Information                    & 9   &                              &    \\
                                            &     & Bait and Switch                       & 1   & Disguised Ads                & 1  
\end{tblr}
}
\end{table*}}

\section{Findings} 
\label{findings}
The dataset ultimately consisted of 148 experimental units from 27 papers, published from 2019 to 2025. A summary of the dataset is reported in Table~\ref{tab:dataset_summary}, which shows the most common recruiting methods, distribution of experimental units per paper, and the distribution in number of participants included for each experiment. 
See Appendix~\ref{sec:supptable}~Table~\ref{units_per_paper} for a full list of the 27 papers that made the final cut and how many experimental units are included from each paper.

\subsection{DMP Experiments Have Covered All High-Level DMP Types}
In total, the 148 experiments tested the effects for a range of DMP types. Table~\ref{tab:dp_breakdown} shows the breakdown of the DMPs in the dataset of experiments, at the high-, meso-, and low-levels~\cite{gray2024ontology}. 
All five high-level patterns appear in the dataset but to uneven degrees. \textsc{Interface Interference} was the most commonly tested high-level pattern, showing up 157 times in the dataset ($N=85/148$ exp. units; 22/27 papers)---most of which are \textsc{Visual Prominence} or \textsc{False Hierarchy}. 76 instances of \textsc{Social Engineering} were tested ($N=45/148$ e.; 6/27 p.), followed by 45 instances of \textsc{Obstruction} ($N=41/148$ e.; 10/27 p.), and 21 instances of \textsc{Forced Action} ($N=21/148$ e.; 6/22 p.). Only 10 instances of \textsc{Sneaking} were tested in the dataset ($N=10/148$ e.; 4/27 p.).

\begin{figure*}
    \centering
    \begin{minipage}[t]{0.49\textwidth}
        \centering
        \includegraphics[width=\linewidth]{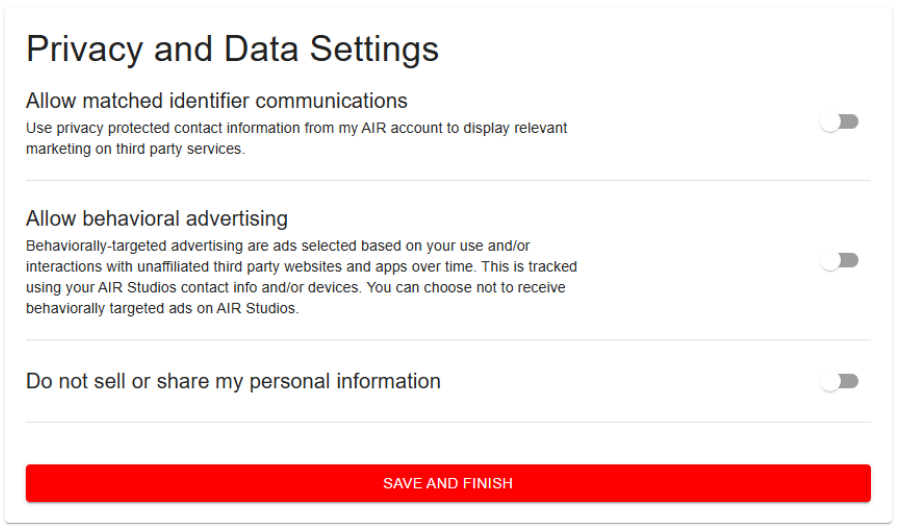}
        \caption*{(a)}
    \end{minipage}
        \begin{minipage}[t]{0.50\textwidth}
        \centering
        \includegraphics[width=\linewidth]{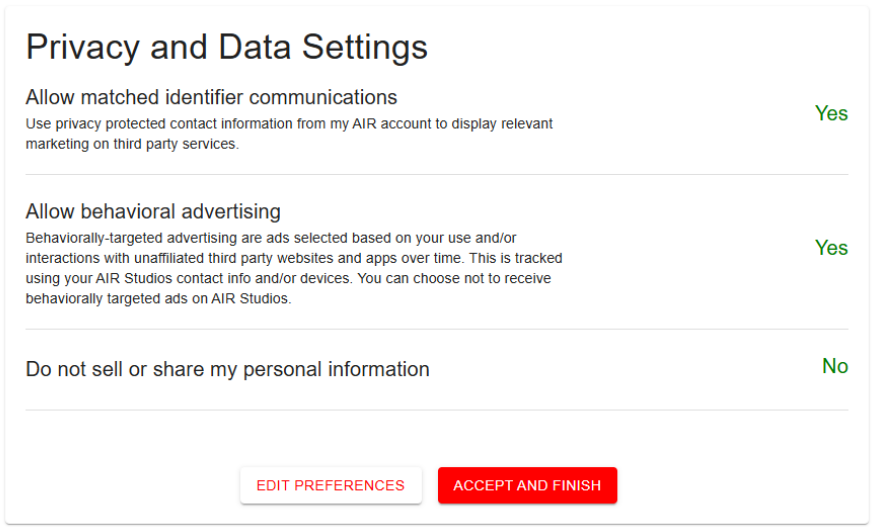}
        \caption*{(b)}
    \end{minipage}
    \hfill
    \caption{Example from a Category 1 experiment testing how the design of privacy settings affected participant decisions while signing up for a video streaming platform. Kugler et al.\ observed a statistically significant increase in participants ``opting'' for data collection when shown an interface with DMPs (b) compared to an interface without (a)~\cite{kugler2025can}.
    The images are adapted from the original work~\cite{kugler2025can}.}
    \label{example_cat1}
    \Description{Shows screenshots from Kugler et al to demonstrate an implementation of a category 1 experiment. Both show different ways of configuring privacy data settings. On the left (a; the control), there are toggles for each of three settings and all are preset to off. At the bottom, there is a single 'save and finish' button. In the right figure (b; the treatment group), the same settings are presented but each is preset to privacy-unfriendly option. Further, instead of a toggle, there is a green ``yes/no'' next to each. For the third setting (do not share my personal information), the green ``no'' is playing on ambiguous language and trick wording. At the bottom, instead of a single save and finish button, there is a 'edit preferences' button and a 'accept all and finish' button. The accept all and finish button is visually more catching.}
\end{figure*}

\subsection{Category 1: DMPs Have a Significant Effect on Participant Behavior}
\label{findings:significance}
Most experiments ($N=101/148$ e.; $23/27$ p.) tested for differences in participant behavior between a control group which was not exposed to DMPs and treatment group(s) that were exposed to a DMP or multiple DMPs, effectively measuring whether DMPs affected behavior to a statistically significant degree (see Figure~\ref{example_cat1} for an example). Of those experiments, a majority ($N=86/101$ e.; 20/27 p.) determined that participant behavior \textit{was changed} when DMPs were experienced, shown by high-level type in Table~\ref{tab:dpvscontrolsummary}. Given the low subdivided sample counts, no pattern type can be claimed to be more likely to result in statistically significant effect---verifiable by pairwise two-tailed \textit{z}-tests for two population proportions.

\aptLtoX{
    \begin{table*}
        \centering
        \caption{All 101 experimental units in Category 1. Each circle represents an experimental unit that tested the effects of DMP(s) compared to a control group by high-level DMP type: Interface Interference (II), Social Engineering (SE), Obstruction (Ob), Forced Action (FA), Sneaking (Sn), or a combination of types (Mult.). \ding{108}$=$statistically significant effects. $\bigcirc$$=$no evidence of significant effects. See \ref{sec:supptable} for a mapping of the paper identifiers.}
        \label{tab:dpvscontrolsummary}
        \Description{Summarizes all category 1 experimental outcomes. There are rows for each publication and columns for high-level dark pattern types. Cells include circles for each experiment conducted by those authors for each high-level dark pattern type. Circles are filled in if the experiment found statistically significant behavioral changes. The final row of the column totals the number of statistically significant results out of total experiments for each high-level type. Vast majority of circles are filled in.}
        \includegraphics[width=0.95\linewidth]{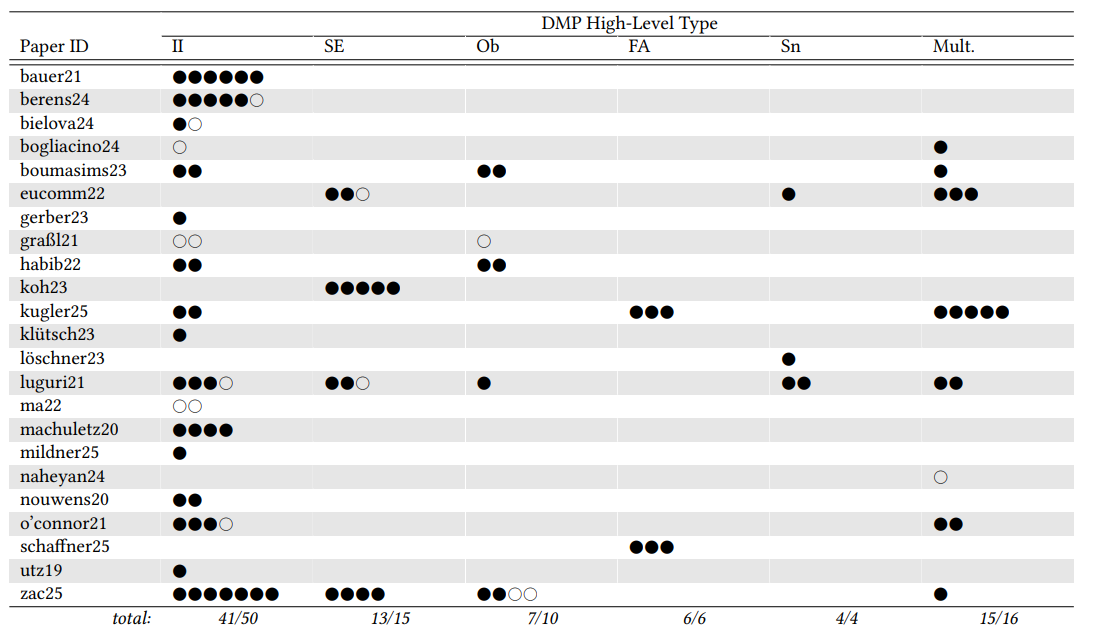}
    \end{table*}

}{
\begin{table*}[t!]
\centering
\caption{All 101 experimental units in Category 1. Each circle represents an experimental unit that tested the effects of DMP(s) compared to a control group by high-level DMP type: Interface Interference (II), Social Engineering (SE), Obstruction (Ob), Forced Action (FA), Sneaking (Sn), or a combination of types (Mult.). \newmoon$=$statistically significant effects. \fullmoon$=$no evidence of significant effects. See \ref{sec:supptable} for a mapping of the paper identifiers.}
\label{tab:dpvscontrolsummary}
\begin{tabular}{>{\hspace{0pt}}m{0.12\linewidth}!{\color{white}\vrule}>{\hspace{0pt}}m{0.12\linewidth}!{\color{white}\vrule}>{\hspace{0pt}}m{0.12\linewidth}!{\color{white}\vrule}>{\hspace{0pt}}m{0.12\linewidth}!{\color{white}\vrule}>{\hspace{0pt}}m{0.12\linewidth}!{\color{white}\vrule}>{\hspace{0pt}}m{0.12\linewidth}!{\color{white}\vrule}>{\hspace{0pt}}m{0.12\linewidth}} 

\hline
                                               & \multicolumn{6}{>{\centering\arraybackslash}m{0.8\linewidth}}{DMP  High-Level Type}  \\ 
\cline{2-7}
                                               Paper ID & II      & SE    & Ob   & FA  & Sn & Mult.                                                                   \\ 
\hline\hline
bauer21                                        & \newmoon\newmoon\newmoon\newmoon\newmoon\newmoon  &       &      &     &   &                                                                         \\
\rowcolor[rgb]{0.90,0.90,0.90} berens24     & \newmoon\newmoon\newmoon\newmoon\newmoon\fullmoon  &       &      &     &    &                                                                         \\
bielova24                                      & \newmoon\fullmoon      &       &      &     &    &                                                                         \\
\rowcolor[rgb]{0.90,0.90,0.90} bogliacino24 & \fullmoon       &       &      &     &    & \newmoon                                                                       \\
boumasims23                                   & \newmoon\newmoon      &       & \newmoon\newmoon   &     &    & \newmoon                                                                       \\
\rowcolor[rgb]{0.90,0.90,0.90} eucomm22     &         & \newmoon\newmoon\fullmoon   &      &     & \newmoon  & \newmoon\newmoon\newmoon                                                                     \\
gerber23                                       & \newmoon       &       &      &     &    &                                                                         \\
\rowcolor[rgb]{0.90,0.90,0.90} graßl21     & \fullmoon\fullmoon      &       & \fullmoon    &     &    &                                                                         \\
habib22                                        & \newmoon\newmoon      &       & \newmoon\newmoon   &     &    &                                                                         \\
\rowcolor[rgb]{0.90,0.90,0.90} koh23        &         & \newmoon\newmoon\newmoon\newmoon\newmoon &      &     &    &                                                                         \\
kugler25                                       & \newmoon\newmoon   &       &      & \newmoon\newmoon\newmoon &    & \newmoon\newmoon\newmoon\newmoon\newmoon                                                                      \\
\rowcolor[rgb]{0.90,0.90,0.90} klütsch23      & \newmoon       &       &      &     &    &                                                                         \\
löschner23                                     &         &       &      &     & \newmoon  &                                                                         \\
\rowcolor[rgb]{0.90,0.90,0.90} luguri21     & \newmoon\newmoon\newmoon\fullmoon    & \newmoon\newmoon\fullmoon   & \newmoon    &     & \newmoon\newmoon & \newmoon\newmoon                                                                      \\
ma22                                           & \fullmoon\fullmoon      &       &      &     &    &                                                                         \\
\rowcolor[rgb]{0.90,0.90,0.90} machuletz20  & \newmoon\newmoon\newmoon\newmoon    &       &      &     &    &                                                                         \\
mildner25                                      & \newmoon       &       &      &     &    &                                                                         \\
\rowcolor[rgb]{0.90,0.90,0.90} naheyan24    &         &       &      &     &    & \fullmoon                                                                       \\
nouwens20                                      & \newmoon\newmoon      &       &      &     &    &                                                                         \\
\rowcolor[rgb]{0.90,0.90,0.90} o'connor21    & \newmoon\newmoon\newmoon\fullmoon    &       &      &     &    & \newmoon\newmoon                                                                      \\
schaffner25                                    &         &       &      & \newmoon\newmoon\newmoon &    &                                                                         \\
\rowcolor[rgb]{0.90,0.90,0.90} utz19        & \newmoon       &       &      &     &    &                                                                         \\
zac25                                          & \newmoon\newmoon\newmoon\newmoon\newmoon\newmoon\newmoon & \newmoon\newmoon\newmoon\newmoon  & \newmoon\newmoon\fullmoon\fullmoon &     &    & \newmoon                                                                       \\
\hline

\multicolumn{1}{r}{\textit{total:}} & \multicolumn{1}{c}{\textit{41/50}} & \multicolumn{1}{c}{\textit{13/15}} & \multicolumn{1}{c}{\textit{7/10}} & \multicolumn{1}{c}{\textit{6/6}} & \multicolumn{1}{c}{\textit{4/4}} & \multicolumn{1}{c}{\textit{15/16}} 
\end{tabular}
\vspace{0.2cm}
\end{table*}
}

\aptLtoX{\begin{figure}
    \centering
        \includegraphics[width=0.6\linewidth]{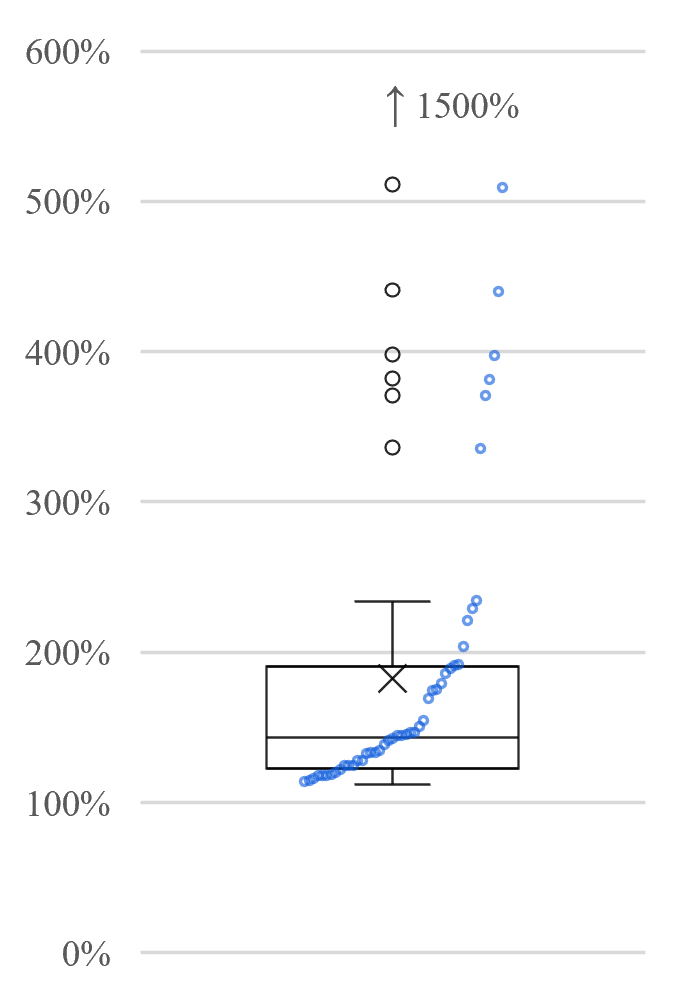}
        \caption{Relative percent increase in the portions of participants experiencing undesirable outcomes in presence of DMPs.}
        \label{fig:relative_box}
        \Description{Shows a distribution of the relative percent increases. A box plot overlays the raw distribution plot.}
    \end{figure}
    \begin{figure}
        \centering
        \includegraphics[width=0.6\linewidth]{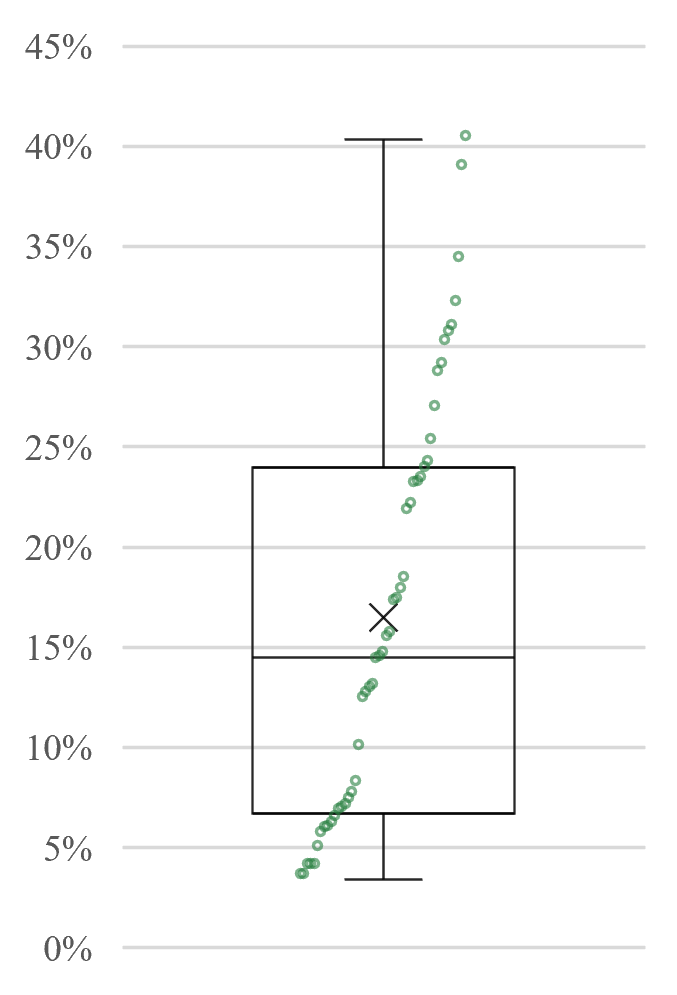}
        \caption{Absolute percent increase in the portions of participants experiencing undesirable outcomes in presence of DMPs.}
        \label{fig:absolute_box}
        \Description{Shows a distribution of the absolute percent increases. A box plot overlays the raw distribution plot.}
\end{figure}}{
\begin{figure*}[tbhp!]
\begin{minipage}[T]{\textwidth}
    \centering
    \begin{minipage}[T]{0.35\textwidth}
        \centering
        \includegraphics[width=\linewidth]{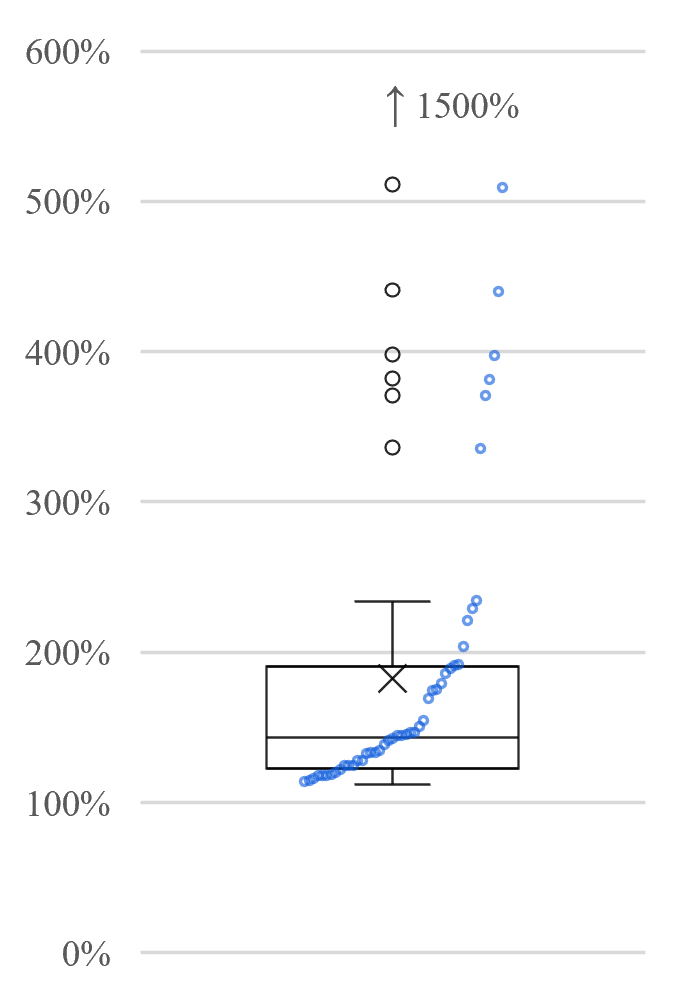}
        \captionof{figure}{\textbf{Relative percent increase} in the portions of participants experiencing undesirable outcomes in presence of DMPs.}
        \label{fig:relative_box}
        \Description{Shows a distribution of the relative percent increases. A box plot overlays the raw distribution plot.}

    \end{minipage}
    \hspace{2.5cm}
    \begin{minipage}[T]{0.35\textwidth}
        \centering
        \includegraphics[width=\linewidth]{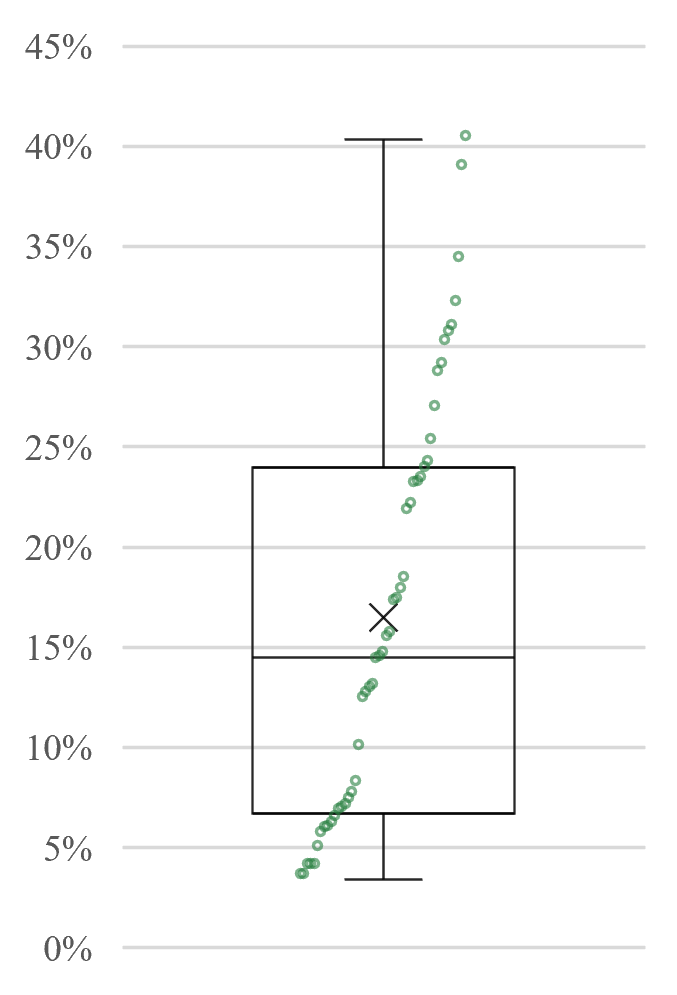}
        \captionof{figure}{\textbf{Absolute percent increase} in the portions of participants experiencing undesirable outcomes in presence of DMPs.}
        \label{fig:absolute_box}
        \Description{Shows a distribution of the absolute percent increases. A box plot overlays the raw distribution plot.}
    \end{minipage}
    \hfill
\end{minipage}
\end{figure*}}

Across all Category 1 experiments, 85\% observed that DMPs had a significant effect on participants.\footnote{The portion of experiments observing significant effects ($86/101 = 85.1\%$) is different from random chance to a high level of significance 
($\alpha=.0005$). The one-tailed one-sample z-test for proportion: $Z = \frac{(p - P)}{\sqrt{\frac{P(1-P)}{n}}} = 7.06 \gg 3.29$, where $p$ is 
 the portion of positive experimental results observed ($0.851$) and $P$ is the expected portion of positive experimental results according to random chance ($0.500$).}
Therefore, while it should be assumed that some degree of publication bias is present,
there remains convincing experimental agreement that DMPs significantly affect participant behavior.

\paragraph{Types of Harms}
The Category 1 experiments that observed significant effects found a variety of harms induced from DMPs.
Most commonly ($N=52/86$ e.; 15/20 p.), participants were found to select privacy un-friendly options as a result of DMPs, such as being more likely to accept unnecessary tracking cookies when pushed by DMPs~\cite[e.g.,][]{habib2022okay}. After privacy-related harms, there is a drop off in the number of studies experimentally measuring other DMP harms. The second most experimentally-measured DMP harm was participants ``opting'' to purchase goods or services they otherwise would not have ($N=29/86$ e.; 3/20 p.)\cite{zac2023dark, luguri2021shining, KOH2023100145}. Other harms included spending more time on a platform than they otherwise would have ($N=3/86$ e.; 1/20 p.)~\cite{schaffner2025experimental}, downloading apps they otherwise would not ($N=1/86$ e.; 1/20 p.)~\cite{lupianez2022behavioural}, and exhibiting reduced ability to distinguish between real news and advertisements ($N=1/86$ e.; 1/20 p.)~\cite{loschner2023different}.

\paragraph{Size of Effects} 
The effects of DMPs were most commonly reported as the difference in the portion of participants experiencing an outcome that they otherwise would not, as concluded by contrast to a control group without the DMPs. For example, Zac et al.\ showed that the use of \textsc{Obstruction} increased the rate at which participants accepted a product promotion from 48.76\% to 61.28\% (p<.001)~\cite{zac2023dark}. 
Of the experiments reporting statistical significance in Category 1, about half reported population means for their control and treatment samples ($N=47/86$ e.; 11/20 p.), allowing for the calculation of comparable effect sizes. 
To control for varied control group means, we calculated both relative and absolute percent increases which are shown in Figure~\ref{fig:relative_box} and Figure~\ref{fig:absolute_box}, respectively. 

Relative percent increases ranged from 112\% to 1500\% (mean = 211\%, median = 144\%). That is, on average, the DMPs tested in these experiments saw participants about 2$\times$ more likely to experience an undesirable outcome---e.g., sharing data they otherwise would not have or purchasing an item they otherwise would not have. 
Absolute percent increases ranged from 3.40\% to 40.4\% (mean = 16.5\%, median = 14.5\%). 
The large variation in effect sizes indicates strong context dependence; that is, the magnitude of DMP effects can vary greatly depending on factors such as the severity of the DMP implemented.
However, we found no correlation between the size of the effect and the type(s) of DMPs being studied or the domain of the experiment.

\begin{figure*}[tbhp!]
    \begin{minipage}[t]{0.46\textwidth}
        \centering
        \raisebox{0.18\height}{
        \fbox{\includegraphics[width=\linewidth]{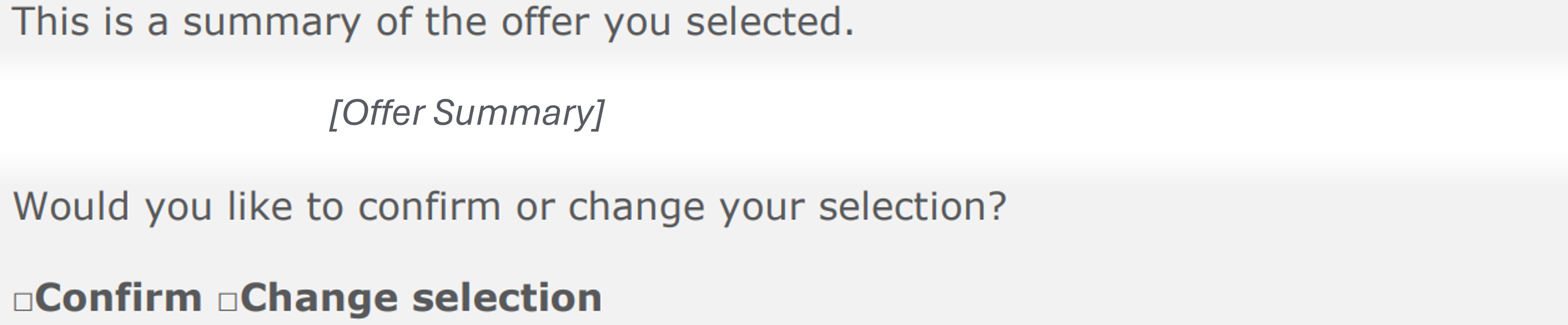}}}
        \caption*{(a)}
    \end{minipage}
    \hspace{0.3cm}
    \begin{minipage}[t]{0.46\textwidth}
        \centering
        \fbox{\includegraphics[width=\linewidth]{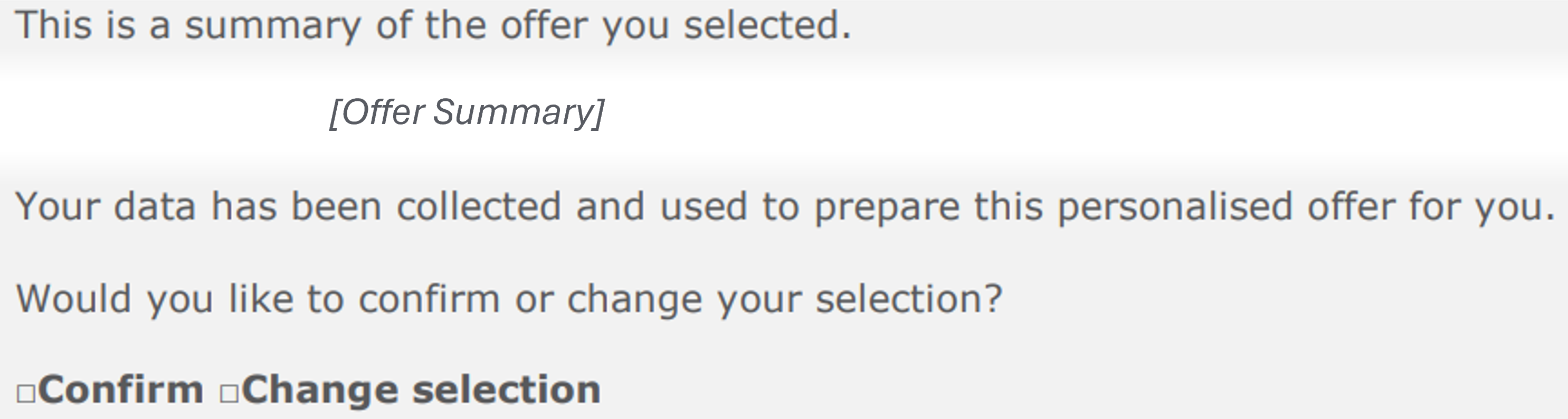}}
        \caption*{(b)}
    \end{minipage}
    \hfill
    \caption{Examples from a Category 2 experiment testing interstitials as interventions. A report by the European Commission tested the effectiveness of two versions of a ``cool down'' interstitial designed to show participants a summary of the decisions they just had made during a subscription sign-up flow that used DMPs. Both versions allowed participants to change their response. Additionally, one version also disclosed to participants the site's use of a \textsc{Social Engineering} DMP (b)~\cite{lupianez2022behavioural}.
    Neither version of the interstitials was found to reduce the effects of the DMPs.
    The images are adapted from the original work~\cite{lupianez2022behavioural}.}
    \label{example_cat2}
    \Description{Shows screenshots from Lupiáñez-Villanueva et al to demonstrate an implementation of a category 2 experiment. On the left (a; the control) there is a popup that includes the text: ``This is a summary of the offer you selected. (Offer summary) Would you like to change your selection? Confirm (checkbox) Change Selection (checkbox).'' On the right (b; the treatment) there is a popup that includes the same text, but also includes ``Your data has been collected and used to prepare a personalized offer for you''}
\end{figure*}

\aptLtoX{
    \begin{table*}
        \centering
        \caption{All experimental units that tested whether external interventions could mitigate the effects of DMPs by high-level DMP type: Interface Interference (II), Social Engineering (SE), Obstruction (Ob), Forced Action (FA), Sneaking (Sn), or a combination of types (Mult.). \ding{108}$=$statistically significant reductions in DMP effects. $\bigcirc$$=$no evidence of significant reductions. See \ref{sec:supptable} for a mapping of the paper identifiers.}
        \label{tab:dpinterventionsummary}
        \Description{Similar to table 4, summarizes all category 2 experimental outcomes. There are rows for each publication and columns for high-level dark pattern types. Cells include circles for each experiment conducted by those authors for each high-level dark pattern type. Circles are filled in if the intervention was found to significant lessen dark pattern effects. The final row of the column totals the number of statistically significant results out of total experiments for each high-level type. This table is much smaller since there are fewer category 2 experiments. Most circles are not filled in.}
        \includegraphics[width=0.6\linewidth]{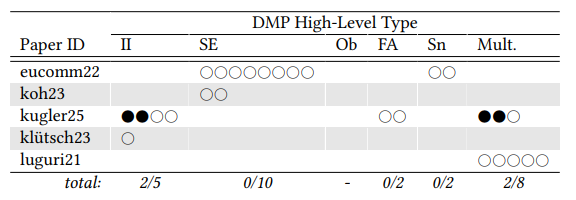}
    \end{table*}

}{

\begin{table*}[hbtp]
\centering

\caption{All experimental units that tested whether external interventions could mitigate the effects of DMPs by high-level DMP type: Interface Interference (II), Social Engineering (SE), Obstruction (Ob), Forced Action (FA), Sneaking (Sn), or a combination of types (Mult.). \newmoon$=$statistically significant reductions in DMP effects. \fullmoon$=$no evidence of significant reductions. See \ref{sec:supptable} for a mapping of the paper identifiers.}
\label{tab:dpinterventionsummary}
\begin{tabular}{l!{\color{white}\vrule}l!{\color{white}\vrule}l!{\color{white}\vrule}l!{\color{white}\vrule}l!{\color{white}\vrule}l!{\color{white}\vrule}l} 
\hline
                                           & \multicolumn{6}{c}{DMP  High-Level Type}                                                                                                                                                               \\ 
\cline{2-7}
                                           Paper ID & II                               & SE                                & Ob                             & FA                               & Sn                               & Mult.~                            \\ 
\hline\hline
eucomm22                                   &                                  & \fullmoon\fullmoon\fullmoon\fullmoon\fullmoon\fullmoon\fullmoon\fullmoon                          &                                &                                  & \fullmoon\fullmoon                               &                                   \\
\rowcolor[rgb]{0.90,0.90,0.90} koh23    &                                  & \fullmoon\fullmoon                                &                                &                                  &                                  &                                   \\
kugler25                                   & \newmoon\newmoon\fullmoon\fullmoon                             &                                   &                                & \fullmoon\fullmoon                               &                                  & \newmoon\newmoon\fullmoon                               \\
\rowcolor[rgb]{0.90,0.90,0.90} klütsch23  & \fullmoon                                &                                   &                                &                                  &                                  &                                   \\
luguri21                                   &                                  &                                   &                                &                                  &                                  & \fullmoon\fullmoon\fullmoon\fullmoon\fullmoon                             \\ 
\hline
\multicolumn{1}{r}{\textit{total:}}                       & \multicolumn{1}{c}{\textit{2/5}} & \multicolumn{1}{c}{\textit{0/10}} & \multicolumn{1}{c}{\textit{-}} & \multicolumn{1}{c}{\textit{0/2}} & \multicolumn{1}{c}{\textit{0/2}} & \multicolumn{1}{c}{\textit{2/8}} 
\end{tabular}
\end{table*}
}

\subsection{Category 2: External Interventions That Could (Not) Mitigate DMP Effects}
\label{sec:findings_interventions}
The next most common experimental type ($N=27/148$ e.; 5/27 p.) were the experiments that tested whether external interventions could counteract the effect of DMPs. External interventions took the form of: 
reflective ``cool down'' interstitials ($N=10/27$ e.; 1/5 p.; see Figure~\ref{example_cat2}, e.g.)~\cite{lupianez2022behavioural}, explicitly priming participants with instructions to select options to maximize privacy ($N=9/27$ e.; 1/5 p.)~\cite{kugler2025can}, 
raising the stakes of participant decisions by increasing the cost of a phony subscription ($N=5/27$ e.; 1/5 p.)~\cite{luguri2021shining}, 
educating participants on DMPs ($N=2/27$ e.; 1/5 p.)~\cite{KOH2023100145}, and educating participants on the consequences of privacy-related decisions ($N=1/27$ e.; 1/5 p.)~\cite{klutsch2023defeating}.

As shown in Table~\ref{tab:dpinterventionsummary} only four experiments ($N=4/27$ e.; all from \cite{kugler2025can}) found evidence that the effect of DMPs could be counteracted by external interventions. 
Specifically, in all four cases, participants were told to maximize their privacy goal with the instructions ``choose the most privacy protective options. [...]'' when signing up for a service, despite any prior privacy beliefs.
Still, the number of experiments ($N=4$) where the intervention counteracted the effect of DMPs is fewer than the experiments (from the same work) finding that the exact same intervention did not significantly reduce the DMP's main effects ($5/9$ e.; also from \cite{kugler2025can}). 
Thus, even when participants were explicitly trying to select what they believed to be the most privacy-friendly options, their efforts were largely unsuccessful. 

Luguri et al.\ tested whether increasing the cost of a subscription pushed by DMPs affected subscription rates. We treat this as an intervention in the form of raising the stakes: by making the financial consequences of participants’ choices more significant. While not an intervention one would propose as a solution to DMPs, the experiments assess whether participants giving heightened effort were still susceptible to DMPs. As the authors discuss, participants facing higher monetary stakes would (in theory) ``be willing to jump over more hurdles in order to save themselves more money'' and ``be less likely to `fall' for the DMPs.''  
However, all experiments testing this intervention ($N=5$; all from~\cite{luguri2021shining}) found that even with cost tripling, DMPs remained just as effective.

There were two versions of the cool down interventions tested (See Figure~\ref{example_cat2}): one that gave participants a summary of their choices ($N=6$) and one that disclosed the use of \textsc{Social Engineering} in addition to the summary ($N=4$)~\cite{lupianez2022behavioural}. Both gave participants the ability to either confirm or change their selections. None of the experiments testing the cool down remedies found significant reductions in the effects of DMPs. 

Further, none of the experiments testing participant education ($N=3$ e.; 2/5 p.) found statistically significant reductions in the effects of DMPs either. 
The tested education strategies included: a 7-minute DMP informational video~\cite{KOH2023100145}, an interactive `spot-the-dark-pattern' activity~\cite{KOH2023100145}, and a virtual-assistant-like pop-up that explained the details of the cookie consent choices participants were exposed to~\cite{klutsch2023defeating}. The researchers of these studies discuss possible reasons for the lack of evidence of effective educational interventions being study limitations, such as participant drop-out~\cite{KOH2023100145}, or cookie decision behavior being already strongly conditioned in Internet users~\cite{klutsch2023defeating}.

\begin{figure*}[t]
    \centering
    \begin{minipage}[t]{0.48\textwidth}
        \centering
        \includegraphics[width=\linewidth]{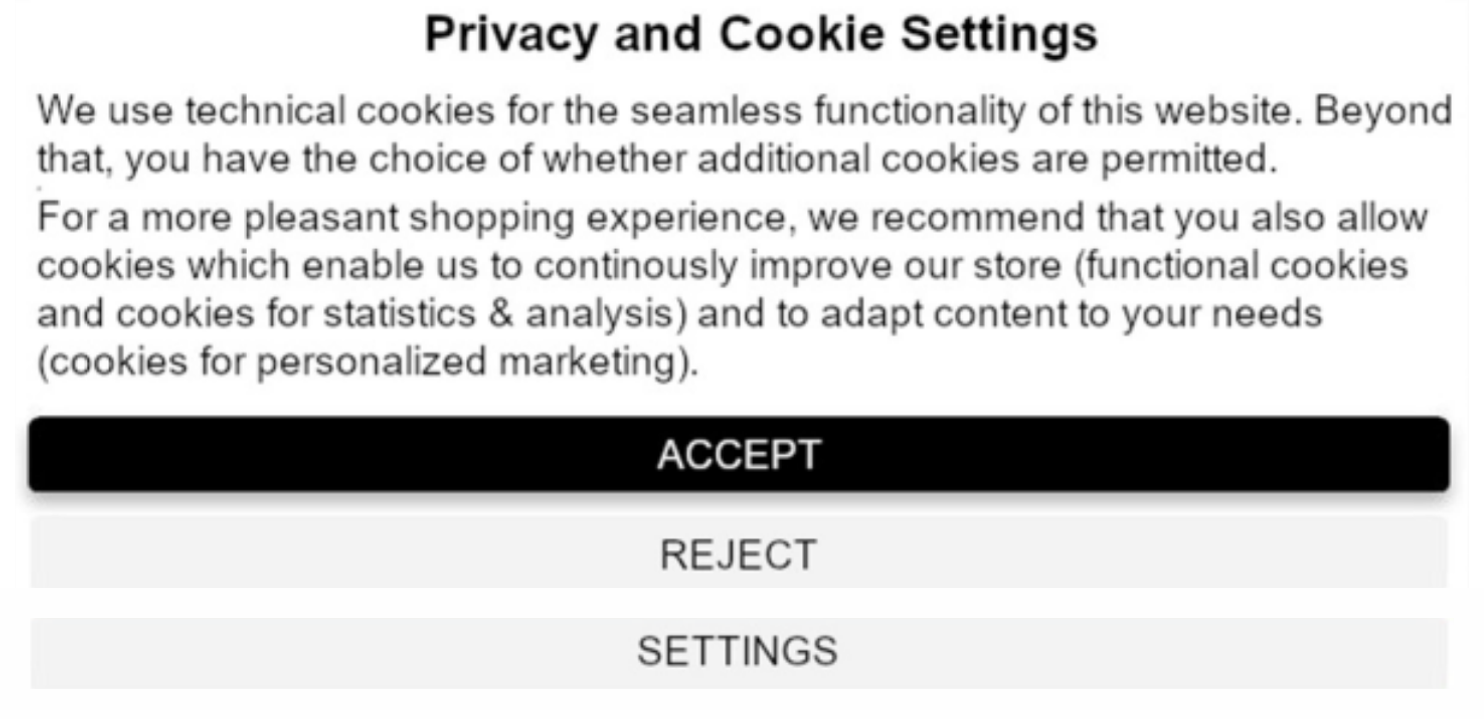}
        \caption*{(a)}
    \end{minipage}
        \begin{minipage}[t]{0.48\textwidth}
        \centering
        \raisebox{0.3cm}{\includegraphics[width=\linewidth]{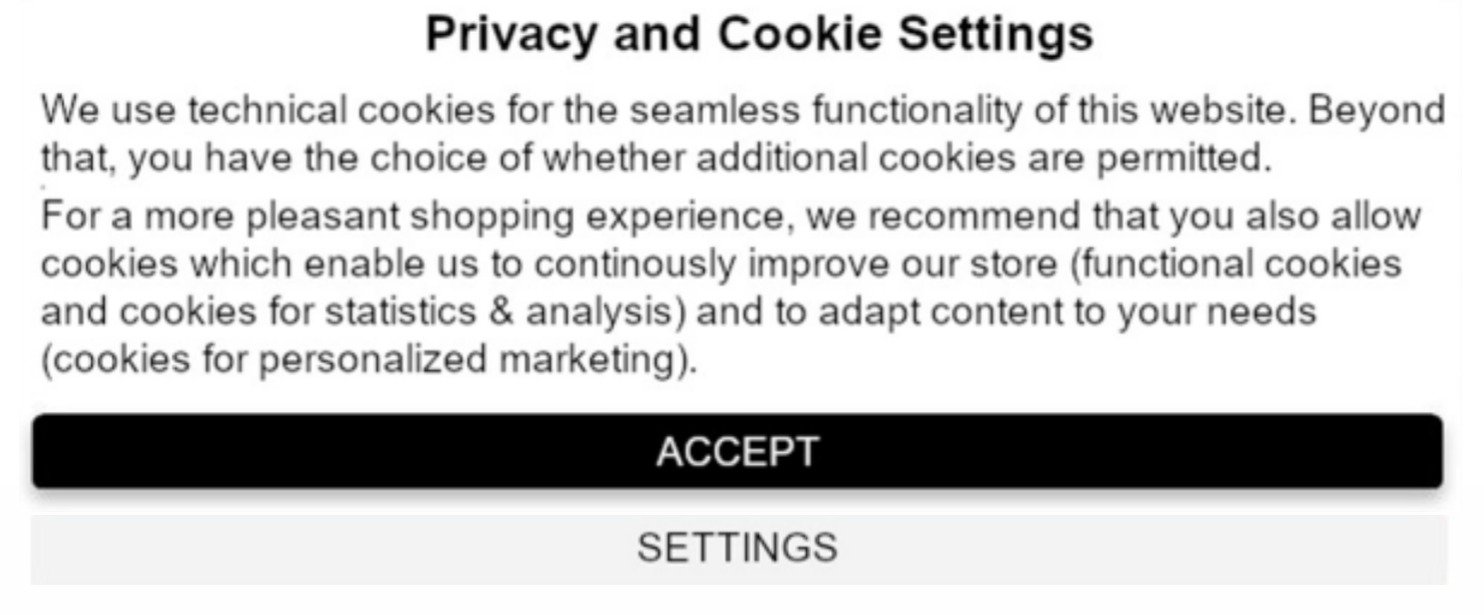}}
        \caption*{(b)}
    \end{minipage}
    \hfill
    \caption{Example of a Category 3 experiment. Mager et al.\ tested the difference between two cookie consent interfaces, one with just \textsc{Interface Interference} (a) and one with the same \textsc{Interface Interference} as well as added \textsc{Obstruction} (b)~\cite{mager2021effectiveness}. They found a statistically significant difference in user consent between the two versions. The images are adapted from the original work~\cite{mager2021effectiveness}.}
    \label{example_cat3}
    \Description{Shows screenshots from Mager et al to demonstrate an implementation of a category 3 experiment. On the left (a) shows a cookie consent interface. After the standard cookie explanation, there are three options: Accept, Reject, and Settings. The accept button is visually most catching. On the right (b) shows the same cookie interface but there is only Accept and Reject, not Settings.}
\end{figure*}

\begin{figure*}[hb!tp]
\centering
    \begin{minipage}[t]{0.44\textwidth}
        \centering
        \includegraphics[width=\linewidth]{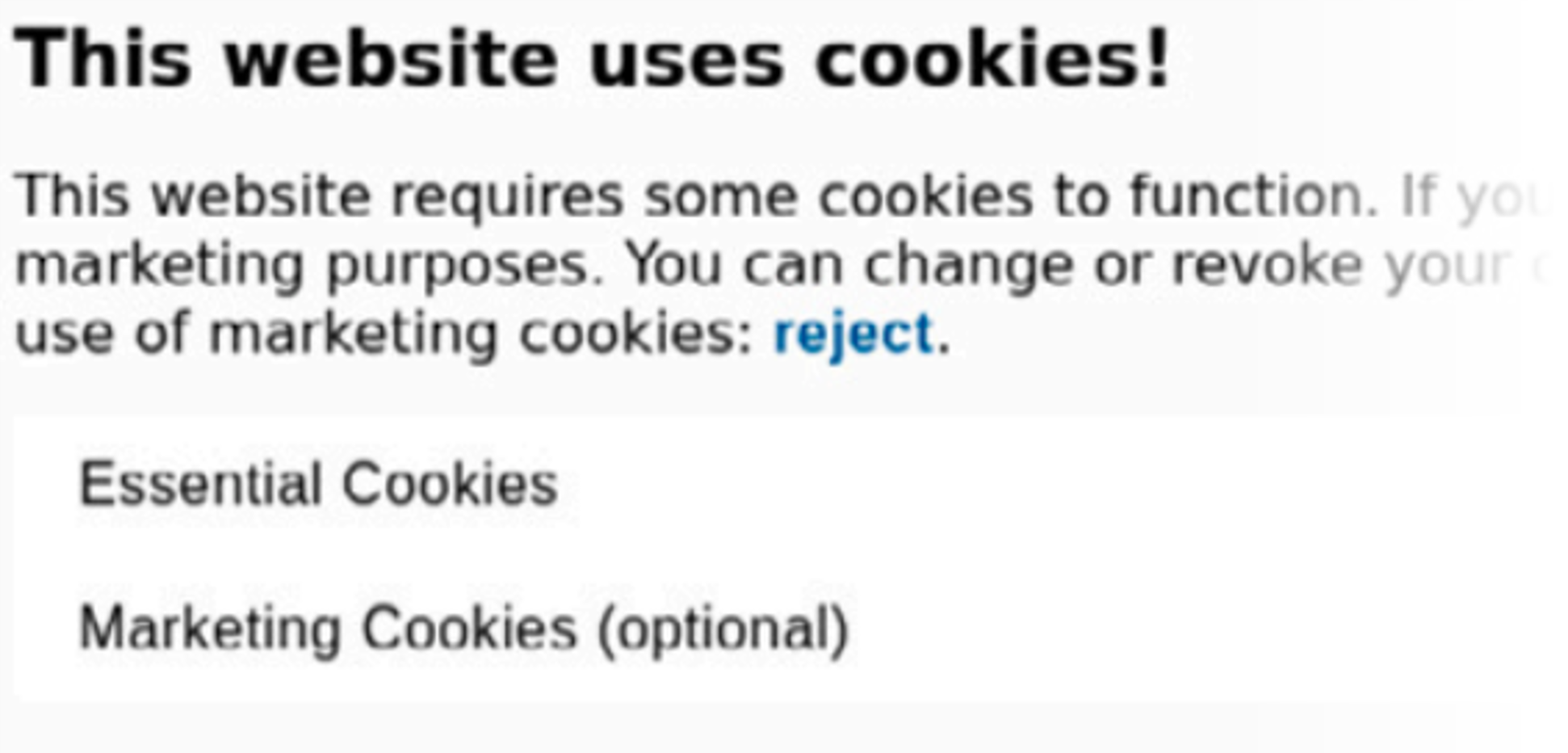}
        \caption*{(a)}
    \end{minipage}
    \begin{minipage}[t]{0.44\textwidth}
        \centering
        \includegraphics[width=\linewidth]{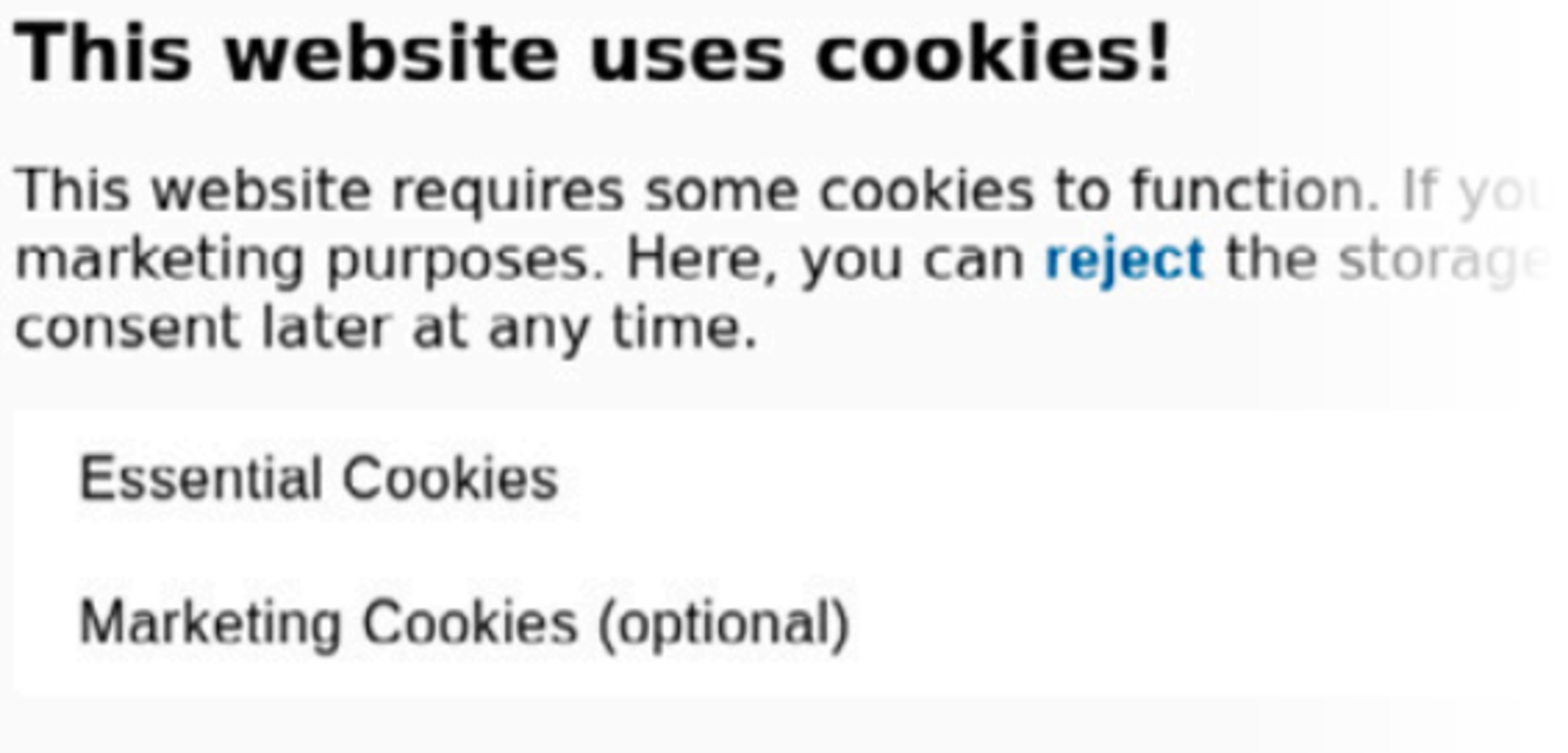}
        \caption*{(b)}
    \end{minipage}
    \hfill
    \caption{Example of a Category 4 experiment. Berens et al.\ tested whether embedding a cookie opt-out hyperlink at the end (a) or middle (b) of a consent notice---two distinct instances of \textsc{Interface Interference}---affected participant consent rates~\cite{berens2024cookie}. They found no evidence of divergent behavior between the two groups. The images are adapted from the original work~\cite{berens2024cookie}.}
    \label{example_cat4}
    \Description{Shows screenshots from Berens et al to demonstrate an implementation of a category 4 experiment. There is another implementation of a cookie consent interface. Both have options to reject cookies as hyperlink embedded in the cookie description. On the left (a), the reject option comes at the end of the description, but on the right (b), the option comes in the middle.}

\end{figure*}

\subsection{Category 3: Whether DMPs Have Additive Effects When Layered is Inconclusive}
\label{sec:findings_stacking}
Some experiments ($N=11/148$ e.; 5/27 p.) tested the effects comparing two conditions of DMPs where one group experienced a subset of the DMPs of a different group, effectively measuring the marginal effect of additional DMPs (See Figure~\ref{example_cat3} for an example).

Category 3 experiments form a small subset of the overall literature reviewed and the strength and generalizability of these findings is relatively limited. 
Three of these studies ($N=4/11$ e.; 3/5 p.)~\cite{mager2021effectiveness, 10.1145/3476087, bogliacino2024testing} found evidence that adding an additional DMP on top of existing DMPs further changed participant behavior. 
The other studies ($N=7/11$ e.; 2/5 p.)~\cite{kugler2025can, lupianez2022behavioural} found no evidence of increased effect from additional DMP exposure. 
Thus, while DMPs do effectively change participant behavior, more research is needed to confirm whether they have diminishing marginal effects when multiple are used in conjunction.

\aptLtoX{
    \begin{table*}
        \centering
        \caption{All papers that examined the potential mediating role of personal characteristics on the effects of DMP. \ding{108}$=$significant correlation between all measures of the attribute and the strength of the DMP effect. $\bigcirc$$=$no evidence of attribute interaction with DMPs for all measures. \LEFTcircle$=$some but not all measures indicated a correlation. $^{+}=$ a positive correlation with DMP effect strength. $^{-}=$ a negative correlation with DMP effect strength. Some papers found correlations in both directions depending on the measure used ($^{+/-}$).}
        \label{tab:personchar_summary}
        \Description{Summarizes all category 5 experimental outcomes. There are rows for each publication and columns for personal characteristics. Cells include circles for each paper that conducted correlation testing between personal characteristics and dark pattern effects. Circles are filled in if the authors found correlations on all measures, unfilled if they found no evidence of correlation on any measure, and half filled if they found correlations on some but not all measures. The filled (or partly-filled) circles are also decorated with +/- signs to indicate the directions of the correlations.}
        \includegraphics[width=0.95\linewidth]{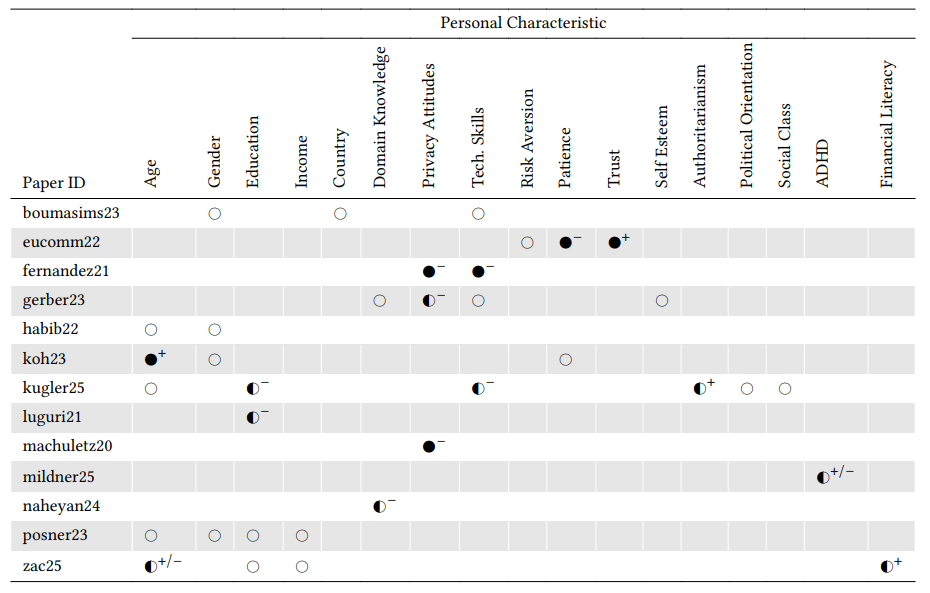}
    \end{table*}

}{
\definecolor{Alto}{rgb}{0.90,0.90,0.90}
\begin{table*}[tb!]
\centering
\caption{All papers that examined the potential mediating role of personal characteristics on the effects of DMP. \newmoon$=$significant correlation between all measures of the attribute and the strength of the DMP effect. \fullmoon$=$no evidence of attribute interaction with DMPs for all measures. \LEFTcircle$=$some but not all measures indicated a correlation.  $^{+}=$ a positive correlation with DMP effect strength. $^{-}=$ a negative correlation with DMP effect strength. Some papers found correlations in both directions depending on the measure used ($^{+/-}$).}
\label{tab:personchar_summary}
\resizebox{0.90\linewidth}{!}{
\begin{tblr}{
  row{4} = {Alto},
  row{6} = {Alto},
  row{8} = {Alto},
  row{10} = {Alto},
  row{12} = {Alto},
  row{14} = {Alto},
  cell{1}{2} = {c=17}{c},
  hline{1,3,16} = {-}{},
  hline{2} = {2-18}{},
  vlines = {white},
}
                               & Personal Characteristic           &                                      &                                         &                                      &                                       &                                                &                                                 &                                            &                                             &                                        &                                     &                                           &                                                &                                                     &                                            &                                    &                                                  \\ Paper ID
\begin{sideways}\end{sideways} & \begin{sideways}Age\end{sideways} & \begin{sideways}Gender\end{sideways} & \begin{sideways}Education\end{sideways} & \begin{sideways}Income\end{sideways} & \begin{sideways}Country\end{sideways} & \begin{sideways}Domain Knowledge\end{sideways} & \begin{sideways}Privacy Attitudes\end{sideways} & \begin{sideways}Tech. Skills\end{sideways} & \begin{sideways}Risk Aversion\end{sideways} & \begin{sideways}Patience\end{sideways} & \begin{sideways}Trust\end{sideways} & \begin{sideways}Self Esteem\end{sideways} & \begin{sideways}Authoritarianism\end{sideways} & \begin{sideways}Political Orientation\end{sideways} & \begin{sideways}Social Class\end{sideways} & \begin{sideways}ADHD\end{sideways} & \begin{sideways}Financial Literacy\end{sideways} \\
boumasims23                   &                                   &        \fullmoon                              &                                         &                                      &                     \fullmoon                  &                                                &                                                 &            \fullmoon                                &                                  &                        &                    &                                           &                                                &                                                     &                                            &                                    &                                                  \\
eucomm22                    &                                   &                            &                                         &                                      &                             &                                                &                                                 &                                &           \fullmoon                                    &            \newmoon$^{-}$                             &       \newmoon$^{+}$                                &                                           &                                                &                                                     &                                            &                                    &                                                  \\
fernandez21                    &                                   &                                      &                                         &                                      &                                       &                                                & \newmoon$^{-}$                                 & \newmoon$^{-}$                            &                                             &                                        &                                     &                                           &                                                &                                                     &                                            &                                    &                                                  \\
gerber23                       &                                   &                                      &                                         &                                      &                                       & \fullmoon                                      & \LEFTcircle$^{-}$                                & \fullmoon                                  &                                             &                                        &                                     & \fullmoon                                 &                                                &                                                     &                                            &                                    &                                                  \\
habib22                        & \fullmoon                         & \fullmoon                            &                                         &                                      &                                       &                                                &                                                 &                                            &                                             &                                        &                                     &                                           &                                                &                                                     &                                            &                                    &                                                  \\
koh23                          & \newmoon$^{+}$                   & \fullmoon                            &                                         &                                      &                                       &                                                &                                                 &                                            &                                             & \fullmoon                              &                                     &                                           &                                                &                                                     &                                            &                                    &                                                  \\
kugler25                       & \fullmoon                         &                                      & \LEFTcircle$^{-}$                        &                                      &                                       &                                                &                                                 & \LEFTcircle$^{-}$                           &                                             &                                        &                                     &                                           & \LEFTcircle$^{+}$                               & \fullmoon                                           & \fullmoon                                  &                                    &                                                  \\
luguri21                       &                                   &                                      & \LEFTcircle$^{-}$                        &                                      &                                       &                                                &                                                 &                                            &                                             &                                        &                                     &                                           &                                                &                                                     &                                            &                                    &                                                  \\
machuletz20                    &                                   &                                      &                                         &                                      &                                       &                                                & \newmoon$^{-}$                                 &                                            &                                             &                                        &                                     &                                           &                                                &                                                     &                                            &                                    &                                                  \\
mildner25                      &                                   &                                      &                                         &                                      &                                       &                                                &                                                 &                                            &                                             &                                        &                                     &                                           &                                                &                                                     &                                            & \LEFTcircle$^{+/-}$                 &                                                  \\
naheyan24                      &                                   &                                      &                                         &                                      &                                       & \LEFTcircle$^{-}$                               &                                                 &                                            &                                             &                                        &                                     &                                           &                                                &                                                     &                                            &                                    &                                                  \\
posner23                       & \fullmoon                         & \fullmoon                            & \fullmoon                               & \fullmoon                            &                                       &                                                &                                                 &                                            &                                             &                                        &                                     &                                           &                                                &                                                     &                                            &                                    &                                                  \\
zac25                          & \LEFTcircle$^{+/-}$                &                                      & \fullmoon                               & \fullmoon                            &                                       &                                                &                                                 &                                            &                                             &                                        &                                     &                                           &                                                &                                                     &                                            &                                    & \LEFTcircle$^{+}$                                 
\end{tblr}
}
\end{table*}}

\subsection{Category 4: Comparisons Between DMP Types are Implementation Dependent}
\label{sec:findings_dpvsdp}
The remaining experiments ($N=9/148$ e.; 6/27 p.) tested behavioral differences between multiple experimental conditions where one cannot be considered a subset of another, effectively measuring the comparative effect between different DMPs or sets of DMPs (See Figure~\ref{example_cat4} for an example). 
Results of these experiments were mixed and fail to provide overall conclusions on the comparative effects of DMPs.
As with Category 3 experiments, these experiments make up a small subset of the overall literature reviewed. As such, findings in this section should be weighed accordingly.

Some Category 4 experiments ($N=6/9$ e.; 5/6 p.) showed that certain DMP instances were more effective than others.
Results of these experiments showed: (i) that certain instances of \textsc{Interface Interference} and \textsc{Obstruction} more effectively increased participant data sharing than others ($N=3/6$ e.; 2/5 p.)~\cite{berens2024cookie, sernac},
(ii) \textsc{Limited Time Messaging} was more effective at guiding participant purchase decisions than three other \textsc{Social Engineering} DMP types ($N=1/6$ e.; 1/5 p.)~\cite{KOH2023100145}, 
(iii) \textsc{Negative Framing} more strongly resulted in participant's privacy-unfriendly behavior than \textsc{Positive Framing} ($N=1/6$ e.; 1/5 p.)~\cite{ma2022prospective}, and
(iv) the change of one instance of \textsc{Bad Defaults} to separate, more severe instances had a significant effect on participant erroneous spending ($N=1/6$ e.; 1/5 p.)~\cite{posner2023dark}.\footnote{While the authors call this a single DMP---``Dark Defaults'', referring to the change itself, rather than the comparison of two different instances of DMPs---, such a temporal interpretation does not currently fit into the leading DMPs ontology.} 

However, Category 4 experimental results were not always significant ($N=3/9$ e.; 2/6 p.). 
Two experiments found that participants experiencing varied versions of
\textsc{Forced Action} and \textsc{Interface Interference}
did not statistically differ from one another~\cite{kugler2025can}. 
Lastly, one experiment found that slightly different instances of the same DMP were found to have no statistical difference~\cite{berens2024cookie}. 
Ultimately, more experiments need to be conducted in order to make conclusive claims about which DMPs are more likely to cause harm. 
However, results from these initial studies indicate that specific implementation details (such as DMP complexity, severity, timing, etc.) may be more important than DMP type alone in predicting harm.

\subsection{Category 5: The Role of Personal Characteristics}
\label{findings:personalchar}
A number of papers (13/27) reported testing for correlations between the effects of DMPs and users' personal characteristics---ranging from demographics (e.g., age and gender) to personal attributes (e.g., technology affinity and political affiliation).
These are studies that test whether the presence of a personal characteristic affected the likelihood or strength of a DMP's effect.
For example, after Lupiáñez-Villanueva et al.\ observed how DMPs led consumers to make choices that they would not have made otherwise (recall Table~\ref{tab:dpvscontrolsummary}), they further tested whether the effects of DMPs were mediated by three self-reported personal characteristics~\cite{lupianez2022behavioural}.
They found that the effects of DMPs were: (i) not correlated with \textit{risk aversion}, (ii) negatively correlated with \textit{patience}, and (iii) positively correlated with \textit{trust}. That is, DMP effects were constant no matter how participants reported their aversion to risk but were higher for participants that reported lower levels of patience or higher levels of trust.
Table~\ref{tab:personchar_summary} summarizes all Category 5 results. As discussed in \S\ref{paper_coding}, this analysis is conducted at the \textit{paper-level} rather than the \textit{experiment-level} since authors' correlation assertions commonly relied on multiple paper-wide measures.

As shown, many correlations are muddied by inconsistency. 
Most (9/15) of the papers that identified correlations between DMP effects and personal characteristics also found lack of correlations (or even correlations in the opposite direction) when using other measures. 
For example, Zac et al.\ found that older participants were more likely to select a product pushed by \textsc{Interface Interference} (a positive correlation) but less likely to select the same product if it were pushed with \textsc{Obstruction} instead (a negative correlation)~\cite{zac2023dark}. 
They also found that age did not interact with participants' willingness to actually complete the payment---beyond just selecting the product---when the product was pushed with \textsc{Obstruction} or \textsc{Social Engineering} (no correlations). 

In summary, experimental evidence that personal characteristics can mediate the effects of DMPs is sparse (but nonetheless present). This supports claims that differences in personal characteristics are less important than the existence or severity of DMPs.
Of course, more research is warranted for conclusivity, especially when considering the specific personal characteristics for which there are still few studies.
However, some attribute interactions stand out as approaching early experimental agreement. For example, of all four works analyzing gender~\cite{bouma2023us, habib2022okay, KOH2023100145, posner2023dark}, none have found significant interactive effects with DMPs. Further, negative correlations between privacy concerns and DMP effects were found in all three papers that tested for it~\cite{10.1145/3476087, gerber2023don, machuletz2019multiple}---but not consistently. 
Additional research is needed to reliably identify which---if any---user groups are consistently more vulnerable to different classes of DMPs.

\section{Discussion}
\label{sec:discussion}

\subsection{Backing Regulatory Discussions with Empirical Agreement}
Given the strong evidence in our review that DMPs result in user harm (\S~\ref{findings:significance}) and the lack of strong evidence that external interventions mitigate the effects of DMPs (\S~\ref{sec:findings_interventions}), we recommend that DMP scholars and policymakers should confront the fact that consumer ``self-help''~\cite{kugler2025can} or other ``aftermarket tools'' may not solve the problem. Even embedded platform tools may not mitigate harms: For example, TikTok's embedded time management tool reduced teens' average daily engagement from about 108.5 minutes by only about 1\%~\cite{allyn2024tiktok}.
This supports the argument that, rather than solely focusing on an increased public awareness of or tolerance for DMPs, the optimal strategy to safeguard consumers from DMPs likely requires removal of DMPs from the digital landscape, a pursuit well-suited for regulatory bodies and standard-setting organizations. 

Regulatory discussions are already underway~\cite{oecd2022dark, eu_edpb_dps, ftc_dpreport_2022}. Our review adds weight to these discussions by uncovering agreement within DMPs experimental scholarship regarding their significant effects and resilience to interventions, while recognizing that certain DMPs, especially \textsc{Forced Action} and \textsc{Sneaking}, have not been studied as thoroughly.

\subsection{Untangling Tensions and Addressing Experimental Gaps}
Despite the number of experiments we reviewed and the conclusions reached that DMPs commonly have an effect on consumer behavior, we find that there remain some important open questions.

For one, the literature offers only preliminary results on two key issues: (1) the relative strengths of different DMP types (\S~\ref{sec:findings_dpvsdp}), and (2) the extent to which combining (``stacking'') multiple DMPs amplifies consumer harm (\S~\ref{sec:findings_stacking}). We recommend future experimental work address these gaps. 

There are also dimensions of DMPs that we observed to be overlooked by experimentalists. 
Most experiments thus far focus on privacy settings, followed by e-commerce. Other possible harms put forth by Mathur et al.'s landmark DMPs work~\cite{mathur2021makes} include harms to individual autonomy and societal welfare.
Such non-material harms (as Santos et al.\ call them~\cite{santos2025no}) related to attention, autonomy, and critical thinking are quite under-represented in the corpus of conducted experiments. 
While there has been some progress in attempting to isolate and measure societal harms from DMPs~\cite{posner2023dark}---such as their impact on fair markets or cultural perceptions---much work remains to be done, as these effects are still not fully understood and remain difficult to measure in controlled settings.

Other open questions involve the dominant temporal lens of most experimental work.
The bulk of studies address short-term, immediate behavioral responses to isolated DMP exposures. 
However, emerging scholarship suggests that DMPs may have effects that persist or even compound over time and repeated interactions. 
For instance, sudden changes to an interface can be used to exploit inertia or muscle memory in repeat users~\cite{10.1145/3605655.3605664}, and exposure to DMPs in a cookie interfaces can change participant behavior on future, non-dark-pattern cookie interfaces~\cite{bielova2024effect}. 
We did not identify any other work that tested how the effects of DMPs may last beyond the interaction at-hand, potentially indicating an insufficiency in existing experimental approaches for capturing consequences across extended timescales. 
We encourage researchers to examine long-term and non-material harms of DMPs, as these remain underexplored. Closing these gaps will benefit both future studies and effective policy-making.

\subsection{Evidentiary Standards and the Role of Experiments} \label{sec:evidence}
Understanding the evidentiary basis of DMP research is crucial, especially as regulatory and legal bodies often rely on scientific findings. Recent work has analyzed the range of research methodologies used specifically in the DMPs literature, detailing the advantages and limitations of each~\cite{gunawan-2025-ipr}. Controlled experiments are typically lauded for their ability to demonstrate causality and offer quantifiable measures of behavioral impact, but they often trade off ecological validity and can be cost prohibitive at scale. In contrast, methods such as observational studies, qualitative content analysis, and surveys can provide broader contextual understanding and surface non-obvious occurrences.

Notably, legal and regulatory decisions around DMPs have historically drawn on a variety of evidence types~\cite{gunawan-2025-ipr}, 
including non-experimental methods like expert evaluations and telemetry data analysis.
Thus, experiments are not the sole source of actionable evidence; policy and enforcement have often relied on a triangulation of methods.
In sum, developing a robust, actionable understanding of DMP harms requires integrating the rigor of experimental studies with the breadth of insights from observational and qualitative research.

\section{Conclusion}
\label{sec:conclusion}
Through a systematic literature review, our study puts forth that there is compelling field-wide evidence that DMPs significantly alter user behavior.
Studies included in our analysis predominantly indicate that DMPs lead to undesirable outcomes, such as privacy invasions or avoidable purchases.
Further, our review indicates that, despite some efforts to counteract the effects of DMPs through external interventions like privacy goal-setting and user education, DMPs remain effective---pointing to the need for regulatory efforts that remove DMPs altogether rather than relying on consumer education. This is further supported by the notion that DMPs seem to similarly affect all users. 
Our review also underscores the need for more experiments to be conducted before concluding whether DMPs have a compounding effect or whether certain DMPs are more pernicious than others. 
Gaps in DMP research remain, such as expanding experimental results for less-studied harms and understanding DMPs long-term or society-wide impacts. 
Given that DMPs have significant effects on user behavior, we urge HCI researchers and policymakers to build on this foundation to address both the immediate harms and the persistent, evolving impact of DMPs in the digital landscape.

\bibliographystyle{ACM-Reference-Format}
\bibliography{references}

\appendix

\onecolumn

\section{Supplementary Materials}

\subsection{Search Queries}
\label{sec:queries}
Google Scholar and ACM DL query: (``dark pattern'' OR ``dark patterns'' OR ``dark design'' OR ``dark designs'' OR ``deceptive design'' OR ``deceptive designs'' OR ``manipulative design'' OR ``manipulative designs'' OR ``coercive design'' OR ``coercive designs'' OR ``dark default'' OR ``dark defaults'') AND (experiment*). SSRN query: ``dark patterns OR dark defaults''. ArXiv query ``dark patterns''. 

\subsection{Supplementary Tables}
\label{sec:supptable}

\aptLtoX{
  \begin{table}[tb!]
  \centering
  \caption{List of included papers and the number of experimental units per paper. The government reports are bolded.}
  \label{units_per_paper}

  \resizebox{\linewidth}{!}{
  \begin{tabular}{p{3cm} p{1.5cm} p{12cm}}
  \hline
  \textbf{Paper ID} & \textbf{Exp.\ Units} & \textbf{Citation} \\
  \hline
  \rowcolor{Alto}
  bauer21~\cite{bauer2021you} & 6 &
  \textcolor{Shark}{Bauer, J. M., Bergstrøm, R., Foss-Madsen, R. (2021). Are you sure, you want a cookie?–The effects of choice architecture on users' decisions about sharing private online data. \textit{Computers in Human Behavior}, 120, 106729.} \\

  berens24~\cite{berens2024cookie} & 8 &
  \textcolor{Shark}{Berens, B. M., Bohlender, M., Dietmann, H., Krisam, C., Kulyk, O., Volkamer, M. (2024). Cookie disclaimers: Dark patterns and lack of transparency. \textit{Computers \& Security}, 136, 103507.} \\

  \rowcolor{Alto}
  bielova24~\cite{bielova2024effect} & 2 &
  \textcolor{Shark}{Bielova, N., Litvine, L., Nguyen, A., Chammat, M., Toubiana, V., Hary, E. (2024). The effect of design patterns on (present and future) cookie consent decisions. In \textit{33rd USENIX Security Symposium (USENIX Security 24)} (pp.~2813--2830).} \\

  \textbf{bogliacino24}~\cite{bogliacino2024testing} & 3 &
  \textcolor{Shark}{Bogliacino, F., Pejsachowicz, L., Liva, G., \& Lupiáñez-Villanueva, F. (2023). Testing for manipulation: Experimental evidence on dark patterns. \textit{Available at SSRN 4755295}.} \\

  \rowcolor{Alto}
  boumasims23~\cite{bouma2023us} & 5 &
  \textcolor{Shark}{Bouma-Sims, E. R., Li, M., Lin, Y., Sakura-Lemessy, A., Nisenoff, A., Young, E., Birrell, E., Cranor, L. F., Habib, H. (2023). A US-UK usability evaluation of consent management platform cookie consent interface design on desktop and mobile. In \textit{Proceedings of the 2023 CHI Conference on Human Factors in Computing Systems} (pp.~1--36).} \\

  \textbf{eucomm22}~\cite{lupianez2022behavioural} & 20 &
  \textcolor{Shark}{Lupiáñez-Villanueva, F., Boluda, A., Bogliacino, F., Liva, G., Lechardoy, L., \& de las Heras Ballell, T. R. (2022). Behavioural study on unfair commercial practices in the digital environment: dark patterns and manipulative personalisation. \textit{Publications Office of the European Union}.} \\

  \rowcolor{Alto}
  fernandez21~\cite{10.1145/3476087} & 2 &
  \textcolor{Shark}{Bermejo Fernandez, C., Chatzopoulos, D., Papadopoulos, D., Hui, P. (2021). This website uses nudging: MTurk workers' behaviour on cookie consent notices. \textit{Proceedings of the ACM on Human-Computer Interaction}, 5(CSCW2), 1--22.} \\

  gerber23~\cite{gerber2023don} & 1 &
  \textcolor{Shark}{Gerber, N., Stöver, A., Peschke, J., Zimmermann, V. (2023). Don’t accept all and continue: Exploring nudges for more deliberate interaction with tracking consent notices. \textit{ACM Transactions on Computer-Human Interaction}, 31(1), 1--36.} \\

  \rowcolor{Alto}
  graßl21~\cite{grassl21dark} & 3 &
  \textcolor{Shark}{Graßl, P., Schraffenberger, H., Borgesius, F. Z., Buijzen, M. (2021). Dark and Bright Patterns in Cookie Consent Requests. \textit{Journal of Digital Social Research}, 3(1), 1--38.} \\

  habib22~\cite{habib2022okay} & 4 &
  \textcolor{Shark}{Habib, H., Li, M., Young, E., Cranor, L. (2022). ``Okay, whatever'': An evaluation of cookie consent interfaces. In \textit{Proceedings of the 2022 CHI Conference on Human Factors in Computing Systems} (pp.~1--27).} \\

  \rowcolor{Alto}
  koh23~\cite{KOH2023100145} & 8 &
  \textcolor{Shark}{Koh, W. C., Seah, Y. Z. (2023). Unintended consumption: The effects of four e-commerce dark patterns. \textit{Cleaner and Responsible Consumption}, 11, 100145.} \\

  kugler25~\cite{kugler2025can} & 26 &
  \textcolor{Shark}{Kugler, M. B., Strahilevitz, L., Chetty, M., Mahapatra, C. (2025). Can Consumers Protect Themselves Against Privacy Dark Patterns? \textit{University of Chicago Coase-Sandor Institute for Law \& Economics Research Paper} (25-01).} \\

  \rowcolor{Alto}
  klütsch23~\cite{klutsch2023defeating} & 2 &
  \textcolor{Shark}{Klütsch, J., Böffel, C., von Salm-Hoogstraeten, S., Schlittmeier, S. J. (2023). Defeating Dark Patterns: The impact of supporting information on dark patterns and cookie privacy decisions. \textit{Proceedings TecPsy 2023}, 41.} \\

  löschner23~\cite{loschner2023different}       & 1                   & Löschner, D. M., Pannasch, S. (2023, July). Different ways to deceive: Uncovering the psychological effects of the three dark patterns preselection, confirmshaming and disguised ads. In~\textit{International Conference on Human-Computer Interaction}~(pp. 62-69). Cham: Springer Nature Switzerland.                                                                              \\

  \rowcolor{Alto}
  luguri21~\cite{luguri2021shining}         & 17                  & Luguri, J., Strahilevitz, L. J. (2021). Shining a light on dark patterns.~\textit{Journal of Legal Analysis},~\textit{13}(1), 43-109.                                                                                                                                                                                                                                                  \\

  ma22~\cite{ma2022prospective}             & 3                   & Ma, E., Birrell, E. (2022, April). Prospective consent: The effect of framing on cookie consent decisions. In~\textit{CHI Conference on human factors in computing systems extended abstracts}~(pp. 1-6).                                                                                                                                                                              \\

  \rowcolor{Alto}
  machuletz20~\cite{machuletz2019multiple}      & 4                   & Machuletz, D., Böhme, R. (2020). Multiple Purposes, Multiple Problems: A User Study of Consent Dialogs after GDPR.~\textit{Proceedings on Privacy Enhancing Technologies}.                                                                                                                                                                                                             \\

  mager21~\cite{mager2021effectiveness}          & 1                   & Mager, S., Kranz, J. (2021). On the Effectiveness of Overt and Covert Interventions in Influencing Cookie Consent: Field Experimental Evidence. In~\textit{ICIS}.                                                                                                                                                                                                                      \\

  \rowcolor{Alto}
  mildner25~\cite{mildner2025comparative}         & 1                   & Mildner, T., Fidel, D., Stefanidi, E., Woźniak, P. W., Malaka, R., \& Niess, J. (2025, April). A Comparative Study of How People With and Without ADHD Recognise and Avoid Dark Patterns on Social Media. In~ \textit{Proceedings of the 2025 CHI Conference on Human Factors in Computing Systems}~(pp. 1-17). \\

  naheyan24~\cite{naheyan2024effect}         & 1                   & Naheyan, T., Oyibo, K. (2024, April). The effect of dark patterns and user knowledge on user experience and decision-making. In~\textit{International Conference on Persuasive Technology}~(pp. 190-206).~ \\

  \rowcolor{Alto}
  nouwens20~\cite{10.1145/3313831.3376321}        & 2                   & Nouwens, M., Liccardi, I., Veale, M., Karger, D., Kagal, L. (2020, April). Dark patterns after the GDPR: Scraping consent pop-ups and demonstrating their influence. In~\textit{Proceedings of the 2020 CHI conference on human factors in computing systems}~(pp. 1-13).                                                                                                              \\

  o'connor21~\cite{o2021clear}         & 6                   & O'Connor, S., Nurwono, R., Siebel, A., Birrell, E. (2021, November). (Un) clear and (In) conspicuous: The right to opt-out of sale under CCPA. In~\textit{Proceedings of the 20th Workshop on Workshop on Privacy in the Electronic Society}~(pp. 59-72).                                                                                                                              \\

  \rowcolor{Alto}
  posner23~\cite{posner2023dark}          & 1                   & Posner, N., Simonov, A., Mrkva, K., \& Johnson, E. J. (2023). Dark defaults: How choice architecture steers political campaign donations.~\textit{Proceedings of the National Academy of Sciences}{,~}\textit{120}(40), e2218385120.                                         \\
  schaffner25~\cite{schaffner2025experimental}       & 3                   & Schaffner, B., Ulloa, Y., Sahni, R., Li, J., Cohen, A. K., Messier, N., Gao, L., Chetty, M. (2025). An Experimental Study Of Netflix Use and the Effects of Autoplay on Watching Behaviors. \textit{Proceedings of the ACM on Human-Computer Interaction, 9}(2), 1-22.                                                                                                                                                    \\

  \rowcolor{Alto}
  \textbf{sernac22}~\cite{sernac}         & 1                   & Pavón Mediano, A.~Rabanales F.,~Luengo-Miranda, C. Vergara, G. (2022).~Policy Paper on Cookies Consent Requests: Experimental Evidence of Privacy by Default and Dark Patterns On Consumer Privacy Decision Making. From \textit{Servicio Nacionaldel Consumidor. }                                                                                         \\
  utz19~\cite{10.1145/3319535.3354212}            & 1                   & Utz, C., Degeling, M., Fahl, S., Schaub, F., Holz, T. (2019, November). (Un) informed consent: Studying GDPR consent notices in the field. In~\textit{Proceedings of the 2019 acm sigsac conference on computer and communications security}~(pp. 973-990).                                                                                                                            \\

  \rowcolor{Alto}
  zac25~\cite{zac2023dark}             & 16                  & Zac, A., Huang, Y. C., von Moltke, A., Decker, C., Ezrachi, A. (2023). Dark patterns and consumer vulnerability.~\textit{Behavioural Public Policy}, 1-50.                                                                                                                                                                                                                             

  \hline
  \end{tabular}
  }
  \end{table}
}{
\definecolor{Alto}{rgb}{0.874,0.874,0.874}
\definecolor{Shark}{rgb}{0.125,0.125,0.133}
\definecolor{MineShaft}{rgb}{0.133,0.133,0.133}
\definecolor{Boulder}{rgb}{0.466,0.466,0.466}
\begin{longtblr}[
  caption = {List of included papers and the number of experimental units per paper. The government reports are bolded.},
  label = {units_per_paper},
]{
  width = \linewidth,
  colspec = {Q[150]Q[45]Q[819]},
  row{even} = {Alto},
  cell{2}{3} = {fg=Shark},
  cell{3}{3} = {fg=Shark},
  cell{4}{3} = {fg=Shark},
  cell{5}{3} = {fg=Shark},
  cell{6}{3} = {fg=Shark},
  cell{7}{3} = {fg=Shark},
  cell{8}{3} = {fg=Shark},
  cell{9}{3} = {fg=Shark},
  cell{10}{3} = {fg=Shark},
  cell{11}{3} = {fg=Shark},
  cell{12}{3} = {fg=Shark},
  cell{13}{3} = {fg=Shark},
  cell{14}{3} = {fg=Shark},
  cell{15}{3} = {fg=Shark},
  cell{16}{3} = {fg=Shark},
  cell{17}{3} = {fg=Shark},
  cell{18}{3} = {fg=Shark},
  cell{19}{3} = {fg=Shark},
  cell{20}{3} = {fg=Shark},
  cell{21}{3} = {fg=Shark},
  cell{22}{3} = {fg=Shark},
  cell{23}{3} = {fg=Shark},
  cell{24}{3} = {fg=Shark},
  cell{25}{3} = {fg=Shark},
  cell{26}{3} = {fg=Shark},
  cell{27}{3} = {fg=Shark},
  cell{28}{3} = {fg=Shark},
  hline{1,29} = {-}{},
}
\textbf{Paper ID} & \textbf{Exp. Units} & \textbf{Citation}                                                                                                                                                                                                                                                                                                                                                                      \\
bauer21~\cite{bauer2021you}           & 6                   & Bauer, J. M., Bergstrøm, R., Foss-Madsen, R. (2021). Are you sure, you want a cookie?–The effects of choice architecture on users' decisions about sharing private online data.~\textit{Computers in Human behavior},~\textit{120}, 106729.                                                                                                                                            \\
berens24~\cite{berens2024cookie}          & 8                   & Berens, B. M., Bohlender, M., Dietmann, H., Krisam, C., Kulyk, O., Volkamer, M. (2024). Cookie disclaimers: Dark patterns and lack of transparency.~\textit{Computers Security},~\textit{136}, 103507.                                                                                                                                                                                 \\
bielova24~\cite{bielova2024effect}         & 2                   & Bielova, N., Litvine, L., Nguyen, A., Chammat, M., Toubiana, V., Hary, E. (2024). The effect of design patterns on (present and future) cookie consent decisions. In~\textit{33rd USENIX Security Symposium (USENIX Security 24)}~(pp. 2813-2830).                                                                                                                                     \\
\textbf{bogliacino24}~\cite{bogliacino2024testing}      & 3                   & Bogliacino, F., Pejsachowicz, L., Liva, G., \& Lupiáñez-Villanueva, F. (2023). Testing for manipulation: Experimental evidence on dark patterns.~\textit{Available at SSRN 4755295}.                                                                                                                           \\
boumasims23~\cite{bouma2023us}       & 5                   & Bouma-Sims, E. R., Li, M., Lin, Y., Sakura-Lemessy, A., Nisenoff, A., Young, E., Birrell, E., Cranor, L.F.  Habib, H. (2023, April). A US-UK usability evaluation of consent management platform cookie consent interface design on desktop and mobile. In~\textit{Proceedings of the 2023 CHI Conference on Human Factors in Computing Systems}~(pp. 1-36).                                                  \\
\textbf{eucomm22}~\cite{lupianez2022behavioural}         & 20                  & Lupiáñez-Villanueva, F., Boluda, A., Bogliacino, F., Liva, G., Lechardoy, L., \& de las Heras Ballell, T. R. (2022).~Behavioural study on unfair commercial practices in the digital environment: dark patterns and manipulative personalisation.\textit{Publications Office of the European Union.}          \\
fernandez21~\cite{10.1145/3476087}       & 2                   & Bermejo Fernandez, C., Chatzopoulos, D., Papadopoulos, D., Hui, P. (2021). This website uses nudging: Mturk workers' behaviour on cookie consent notices.~\textit{Proceedings of the ACM on human-computer interaction},~\textit{5}(CSCW2), 1-22.                                                                                                                                      \\
gerber23~\cite{gerber2023don}          & 1                   & Gerber, N., Stöver, A., Peschke, J., Zimmermann, V. (2023). Don’t accept all and continue: Exploring nudges for more deliberate interaction with tracking consent notices.~\textit{ACM Transactions on Computer-Human Interaction},~\textit{31}(1), 1-36.                                                                                                                              \\
graßl21~\cite{grassl21dark}          & 3                   & Graßl, P., Schraffenberger, H., Borgesius, F. Z., Buijzen, M. (2021). Dark and Bright Patterns in Cookie Consent Requests.~\textit{Journal of Digital Social Research},~\textit{3}(1), 1-38.                                                                                                                                                                                           \\
habib22~\cite{habib2022okay}           & 4                   & Habib, H., Li, M., Young, E., Cranor, L. (2022, April). “Okay, whatever”: An evaluation of cookie consent interfaces. In~\textit{Proceedings of the 2022 CHI conference on human factors in computing systems}~(pp. 1-27).                                                                                                                                                             \\
koh23~\cite{KOH2023100145}             & 8                   & Koh, W. C., Seah, Y. Z. (2023). Unintended consumption: The effects of four e-commerce dark patterns.~\textit{Cleaner and Responsible Consumption},~\textit{11}, 100145.                                                                                                                                                                                                               \\
kugler25~\cite{kugler2025can}         & 26                  & Kugler, M. B., Strahilevitz, L., Chetty, M., Mahapatra, C. (2025). Can Consumers Protect Themselves Against Privacy Dark Patterns?.~\textit{University of Chicago Coase-Sandor Institute for Law Economics Research Paper}, (25-01).                                                                                                                                                   \\
klütsch23~\cite{klutsch2023defeating}         & 2                   & Klütsch, J., Böffel, C., von Salm-Hoogstraeten, S., Schlittmeier, S. J. (2023). Defeating Dark Patterns: The impact of supporting information on dark patterns and cookie privacy decisions.~\textit{Proceedings TecPsy 2023}, 41.                                                                                                                                                     \\
löschner23~\cite{loschner2023different}       & 1                   & Löschner, D. M., Pannasch, S. (2023, July). Different ways to deceive: Uncovering the psychological effects of the three dark patterns preselection, confirmshaming and disguised ads. In~\textit{International Conference on Human-Computer Interaction}~(pp. 62-69). Cham: Springer Nature Switzerland.                                                                              \\
luguri21~\cite{luguri2021shining}         & 17                  & Luguri, J., Strahilevitz, L. J. (2021). Shining a light on dark patterns.~\textit{Journal of Legal Analysis},~\textit{13}(1), 43-109.                                                                                                                                                                                                                                                  \\
ma22~\cite{ma2022prospective}             & 3                   & Ma, E., Birrell, E. (2022, April). Prospective consent: The effect of framing on cookie consent decisions. In~\textit{CHI Conference on human factors in computing systems extended abstracts}~(pp. 1-6).                                                                                                                                                                              \\
machuletz20~\cite{machuletz2019multiple}      & 4                   & Machuletz, D., Böhme, R. (2020). Multiple Purposes, Multiple Problems: A User Study of Consent Dialogs after GDPR.~\textit{Proceedings on Privacy Enhancing Technologies}.                                                                                                                                                                                                             \\
mager21~\cite{mager2021effectiveness}          & 1                   & Mager, S., Kranz, J. (2021). On the Effectiveness of Overt and Covert Interventions in Influencing Cookie Consent: Field Experimental Evidence. In~\textit{ICIS}.                                                                                                                                                                                                                      \\
mildner25~\cite{mildner2025comparative}         & 1                   & Mildner, T., Fidel, D., Stefanidi, E., Woźniak, P. W., Malaka, R., \& Niess, J. (2025, April). A Comparative Study of How People With and Without ADHD Recognise and Avoid Dark Patterns on Social Media. In~ \textit{Proceedings of the 2025 CHI Conference on Human Factors in Computing Systems}~(pp. 1-17). \\
naheyan24~\cite{naheyan2024effect}         & 1                   & Naheyan, T., Oyibo, K. (2024, April). The effect of dark patterns and user knowledge on user experience and decision-making. In~\textit{International Conference on Persuasive Technology}~(pp. 190-206).~                                                                                                                                                                             \\
nouwens20~\cite{10.1145/3313831.3376321}        & 2                   & Nouwens, M., Liccardi, I., Veale, M., Karger, D., Kagal, L. (2020, April). Dark patterns after the GDPR: Scraping consent pop-ups and demonstrating their influence. In~\textit{Proceedings of the 2020 CHI conference on human factors in computing systems}~(pp. 1-13).                                                                                                              \\
o'connor21~\cite{o2021clear}         & 6                   & O'Connor, S., Nurwono, R., Siebel, A., Birrell, E. (2021, November). (Un) clear and (In) conspicuous: The right to opt-out of sale under CCPA. In~\textit{Proceedings of the 20th Workshop on Workshop on Privacy in the Electronic Society}~(pp. 59-72).                                                                                                                              \\
posner23~\cite{posner2023dark}          & 1                   & Posner, N., Simonov, A., Mrkva, K., \& Johnson, E. J. (2023). Dark defaults: How choice architecture steers political campaign donations.~\textit{Proceedings of the National Academy of Sciences}{,~}\textit{120}(40), e2218385120.                                         \\
schaffner25~\cite{schaffner2025experimental}       & 3                   & Schaffner, B., Ulloa, Y., Sahni, R., Li, J., Cohen, A. K., Messier, N., Gao, L., Chetty, M. (2025). An Experimental Study Of Netflix Use and the Effects of Autoplay on Watching Behaviors. \textit{Proceedings of the ACM on Human-Computer Interaction, 9}(2), 1-22.                                                                                                                                                    \\
\textbf{sernac22}~\cite{sernac}         & 1                   & Pavón Mediano, A.~Rabanales F.,~Luengo-Miranda, C. Vergara, G. (2022).~Policy Paper on Cookies Consent Requests: Experimental Evidence of Privacy by Default and Dark Patterns On Consumer Privacy Decision Making. From \textit{Servicio Nacionaldel Consumidor. }                                                                                         \\
utz19~\cite{10.1145/3319535.3354212}            & 1                   & Utz, C., Degeling, M., Fahl, S., Schaub, F., Holz, T. (2019, November). (Un) informed consent: Studying GDPR consent notices in the field. In~\textit{Proceedings of the 2019 acm sigsac conference on computer and communications security}~(pp. 973-990).                                                                                                                            \\
zac25~\cite{zac2023dark}             & 16                  & Zac, A., Huang, Y. C., von Moltke, A., Decker, C., Ezrachi, A. (2023). Dark patterns and consumer vulnerability.~\textit{Behavioural Public Policy}, 1-50.                                                                                                                                                                                                                             
\end{longtblr}}

\end{document}